\documentclass[preprint,onecolumn,showpacs,notitlepage]{revtex4-1}

\usepackage{amssymb}
\usepackage{amsmath}
\usepackage{graphicx}
\usepackage{multirow}
\usepackage{caption}
\usepackage{subfig}
\usepackage{float}
\usepackage[autostyle]{csquotes}
\usepackage{color}

\begin{document}

\title
{Cross-correlation Aided Transport in Stochastically Driven Accretion Flows}

\author{Sujit Kumar Nath}
\affiliation{Department of Physics, Indian Institute of Science, Bangalore 560 012, India}
\email{sujitkumar@physics.iisc.ernet.in}

\author{Amit K Chattopadhyay}
\affiliation{Aston University, Nonlinearity and Complexity Research Group, Engineering 
and Applied Science, Birmingham B4 7ET, UK}
\email{a.k.chattopadhyay@aston.ac.uk}


\begin{abstract}

Origin of linear instability resulting in rotating sheared accretion flows has remained a controversial subject for long. While some explanations of such non-normal transient growth of disturbances in the Rayleigh stable limit were available for magnetized accretion flows, similar instabilities in absence of magnetic perturbations remained unexplained. This dichotomy was resolved in two recent publications by Chattopadhyay, {\it et al} where it was shown that such instabilities, especially for non-magnetized accretion flows, were introduced through interaction of the inherent stochastic noise in the system (even a \enquote{cold} accretion flow at 3000K is too \enquote{hot} in the statistical parlance and is capable of inducing strong thermal modes) with the underlying Taylor-Couette flow profiles. Both studies, however, excluded the additional energy influx (or efflux) that could result from nonzero cross-correlation of a noise perturbing the velocity flow, say, with the noise that is driving the vorticity flow (or equivalently the magnetic field and magnetic vorticity flow dynamics). Through the introduction of such a time symmetry violating effect, in this article we show that nonzero noise cross-correlations essentially renormalize the strength of temporal correlations. Apart from an overall boost in the energy rate (both for spatial and temporal correlations, and hence in the ensemble averaged energy spectra), this results in mutual competition in growth rates of affected variables often resulting in suppression of oscillating Alfven waves at small times while leading to faster saturations at relatively longer time scales. The effects are seen to be more pronounced with magnetic field fluxes where the noise cross-correlation magnifies the strength of the field concerned. Another remarkable feature noted specifically for the autocorrelation functions is the removal of energy degeneracy in the temporal profiles of fast growing non-normal modes leading to faster saturation with minimum oscillations. These results, including those presented in the previous two publications, now convincingly explain subcritical transition to turbulence in the linear limit for all possible situations that could now serve as the benchmark for nonlinear stability studies in Keplerian accretion disks.
  
\end{abstract}

\maketitle


\vspace{0.2cm}

\section{Introduction}

Transition to turbulence notwithstanding the linear stability of laminar flow profiles has been the subject of increasing attention in the realm of sheared turbulence \cite{maretzke2014,avila2012,rincon2007,procaccia1983}. While being unconventional in connection with turbulence that is generally associated with growing nonlinear instabilities, such a phenomenology is not altogether unknown, since equivalent predictions in the context of Taylor-Couette twist instability vortices were already made way back in the early nineties \cite{busse1991}. The new surge in interest though is in keeping with experimental projections in relation to the physics of hot accretion flows \cite{gu,kim,mk,rud,dau,zahn,klar,dubrulle_dauchot2005,dubrulle_mari2005,man2005} where rotating shear flows in the Rayleigh stable regime have shown unmistakable signatures of transition to turbulence with a combination of angular momentum increase inspite of decreasing angular speed profiles. Often referred to as \enquote{subcritical turbulence}, especially in the context of Keplerian accretion disks, the subject remained largely controversial since the emergence of such strong power-law driven instabilities leading to huge energy bursts (optimal transient energy $\sim~{\text{Re}}^{2/3}$ \cite{man2005,maretzke2014}) was difficult to explain on the basis of purely hydrodynamic mechanisms. While certain remarkable theoretical headways were made in understanding rotating accretion flow dynamics perturbed by magnetic fluctuations \cite{balbus_hawley1996,balbus_hawley1999,dubrulle_dauchot2005,dubrulle_mari2005}, the interpretation proved grossly insufficient in systems without the presence of such magnetic fluxes. These theories were, however, accurate only within a \enquote{shearing sheet} approximation, a new age acronym for the \enquote{small gap} approximation previously employed \cite{busse1991}, as correctly pointed out \cite{pumir1996,fromang2007}. Such theories also failed to establish any particular critical threshold as a function of relevant system parameters (Reynolds' number, magnetic strength, etc) beyond which such magnetism induced transition to turbulence would take place. Complementary theoretical strategies in absence of magnetic perturbations \cite{man2005} provided some vital clues as to the nature of the dynamical scaling of the energy growth rate. Essentially, it was shown that the presence of non-normal transient eigenmodes were the absolute pre-requisites for explaining such axisymmetric energy bursts in Taylor-Couette flows in presence of a Coriolis force. But once again the system approaches remained reclusive to a simultaneous presence of shear, rotation and magnetic flux. 

This theoretical impasse was broken only recently, as shown in two recent publications by Chattopadhyay {\it et al} \cite{chattopadhyay2013, nath2013}. In these works, the authors introduced noise as a stochastically perturbing (field theoretically) relevant variable alongside the magnetically perturbed Keplerian flows (other types of flow, like constant angular momentum, flat rotation curve and solid body rotation were considered too). Chattopadhyay {\it et al} showed that noise driven hydrodynamic instabilities (vital for cold accretion disks at temperatures of 3000K and above) led to dynamical cross-overs from laminar to turbulent profiles, both in the presence and absence of magnetic perturbations. Close to the linearly stable limit, this led to a new (dynamic) universality class with interesting statistical properties as were detailed in these publications \cite{chattopadhyay2013,nath2013}. The mathematical foundation also accounted for  the large scale (stochastically driven) instabilities in autocorrelation functions in three dimensions, leading to identical instabilities in the energy growth spectrum, revealing large energy growth or even instability in the system as were previously predicted in slightly different contexts \cite{bmraha}.

Starting with the exploration of the effects of a Gaussian distributed white stochastic noise on Rayleigh stable flows (Taylor-Couette flow in presence of Coriolis effects), focusing on the regime governed by transition to turbulence \cite{chattopadhyay2013}, we employed this theoretical architecture to investigate the amplification of linear magnetohydrodynamic 
perturbations in Rayleigh stable rotating, hot, shear flows in the presence of stochastic noise in three dimensions \cite{nath2013}. While \cite{chattopadhyay2013} analyzed a noise driven Rayleigh flow close to the turbulent regime, \cite{nath2013} focused on the effects of noise in the simultaneous presence of a Coriolis perturbed Taylor-Couette flow acted on by a magnetic field. Notably, a case of a linear non-rotating, non-magnetized shear flow was previously studied \cite{ep03} that highlighted a small subset of the more detailed results derived in these works. Both of these works\cite{chattopadhyay2013,nath2013}, including the earlier truncated version \cite{ep03}, though relied on a drastic symmetry assumption related to the nature of the noise distribution. All modes of noise correlations were assumed to be strictly correlated only within their respective configuration hyper spaces, thereby neglecting the effects of cross-correlations that could arise out of sheared cross-sections. The problem with such a major approximation was already shown to be most vital in the understanding of boundary layer sheared flow profiles in the context of regular (non-accretion type) Taylor-Couette flows \cite{akc_shear}. This is not too difficult to envisage either. An approximation of the type essentially implies that the noisy part of (for example) the velocity profile does not, for example, perturb the vorticity or the magnetic flow profiles at all. This might as well be true for a specific parametric regime but there is no ad hoc proof of such a limitation being tenable across the entire parametric space, more importantly in the manifold close to the laminar-turbulent transition regime. The present work extends the scope of the previously laid mathematical foundation even further, now by introducing noise cross-correlations and by analyzing such symmetry violating effects on large (linear) instabilities.

In what follows, section II will outline the basic model where the flow equations would remain identical to \cite{nath2013}, with the only changes being in the noise cross-correlation profiles. Section III would focus on the effects of non-zero noise cross-correlation on temporal (Part A) and spatial (Part B) autocorrelation functions. Section IV too would be subdivided in to two parts: Part A would analyze temporal cross-correlation functions of the variables (velocity ${\bf u}$, vorticity ${\bf \zeta}$, magnetic field ${\bf B}$ and magnetic vorticity ${\bf \zeta_B}$) while Part B would analyze the spatial cross-correlation functions of the same variables. This would be followed up by a summary and discussion in section V. In order to avoid multiplicity of reference, we would allude to the previous published references \cite{chattopadhyay2013,nath2013} as far as is practicable while trying to avoid repetitions. In a slight departure from the previous two works in the series, the present article would not focus on the effects of colored noise, this being already known not to show any qualitative difference in outcomes.

\section{Flow equations: Perturbed magnetized rotating shear flows in presence of cross-correlated noise}

As mentioned already, we would adopt a model identical to \cite{nath2013} and use the same notations in order to retain continuity in discussions.
All quantities would be expressed in dimensionless units where length has been normalized with respect to the system size $L$, time to be measured in units of the inverse of background angular flow velocity $\Omega$ along $z$, velocity to be measured in units of $q\Omega L$ ($1\le q<2$), and other variables expressed as in \cite{man2005,bmraha,chattopadhyay2013,nath2013}). As detailed in \cite{man2005}, $q=3/2$ represents a Keplerian disk, $q=2$ represents a disk with a constant angular momentum while $q=1$ represents a disk with a flat rotation curve.
In such notations and within the ambits of the \enquote{small gap approximation} $-L/2\le x\le L/2$ (\cite{chattopadhyay2013,nath2013}), the background plane shear would be $(0,-x,0)$, the magnetic field would be given by $(0,B_1,1)$, $B_1$ being a constant and the Coriolis angular velocity profile would as previously be defined as
$\Omega\propto r^{-q}$, where $r$ is the radial distance measured. Along with the incompressibility constraint (${\bf \nabla}\cdot{\bf u}=0$) and zero magnetic charge ($\frac{\partial {\bf B}}{\partial t}=0$) imposed on a magnetized Orr-Sommerfeld and Squire flow system that is acted on by a Coriolis force together with a Gaussian distributed white stochastic noise, we get \cite{nath2013}

\begin{subequations}
\begin{equation}
\left(\frac{\partial}{\partial t}-x\frac{\partial}{\partial y}\right)\nabla^2 u
+\frac{2}{q}\frac{\partial \zeta}{\partial z}-\frac{1}{4\pi}\left(B_1\frac{\partial}{\partial y}+\frac{\partial}{\partial z}\right)\nabla^2B_x
=\frac{1}{R_e}\nabla^4 u+\eta_1(x,t),
\label{orrv}
\end{equation}
\begin{equation}
\left(\frac{\partial}{\partial t}-x\frac{\partial}{\partial y}\right)\zeta
+\frac{\partial u}{\partial z}
-\frac{2}{q}\frac{\partial u}{\partial z}-\frac{1}{4\pi}\left(B_1\frac{\partial}{\partial y}+\frac{\partial}{\partial z}\right)\zeta_B=\frac{1}{R_e}\nabla^2 \zeta +
\eta_2(x,t),
\label{zeta}
\end{equation}
\begin{eqnarray}
\left(\frac{\partial }{\partial t}-x\frac{\partial}{\partial y}\right)B_x
-B_1\frac{\partial u}{\partial y}-\frac{\partial u}{\partial z}=
\frac{1}{R_m}\nabla^2B_x+\eta_3(x,t),
\label{orrb}
\end{eqnarray}
\begin{eqnarray}
\left(\frac{\partial }{\partial t}-x\frac{\partial}{\partial y}\right)\zeta_B-
\frac{\partial \zeta}{\partial z}-B_1\frac{\partial \zeta}{\partial y}-
\frac{\partial B_x}{\partial z}=\frac{1}{R_m}\nabla^2\zeta_B+\eta_4(x,t),
\label{orrbzeta}
\end{eqnarray}
\end{subequations}

\noindent
where the velocity and magnetic field perturbations are given by $(u,v,w)$ and $(B_x,B_y,B_z)$ respectively, with
$R_e$ and $R_m$ being the respective hydrodynamic and magnetic Reynolds numbers. $p_{\rm tot}$ is the total
pressure perturbation (including that due to the magnetic field) which has been eleminated from the equations \cite{nath2013} and the most general form of the (colored) noise components $\eta_{1,2,3,4}$ could be given by 

\begin{equation}
<\eta_i({\bf x},t) \eta_j({\bf x'},t')> = D_{ij}({\bf x})\:\delta^3({\bf x}-{\bf x'})\:\delta(t-t').
\end{equation}

\noindent
The asymptotic large distance, long time behavior of the noise statistics have been encapsulated in a pioneering work by Forster, Nelson \& Stephen \cite{nelson}. It has also been previously proved that the effects of finite sized (power law) spatiotemporal correlations would only be vital for nonlinear (stochastically driven) sheared flows \cite{akc_shear} or otherwise for flows affected by multiplicative noise \cite{akc_abhik}. Given that our focal point is a linearly stable Rayleigh flow profile driven by an additive noise, such colored noise statistics would be irrelevant for the universality class under consideration and hence we would restrict ourselves to a white noise itself, that without any loss of generality could be assumed to have the same noise strength for all correlations ($D_{ij}=D_0$).

As discussed in details in \cite{chattopadhyay2013,nath2013}, we would restrict ourselves to the \enquote{small gap} limit \cite{busse1991}, in which, the Fourier expanded flows could be resolved as follows

$u$, $\zeta$, $B_x$, $\zeta_B$ and $\eta_i$ as
\begin{eqnarray}
\nonumber
u({\bf x},t)=\int\tilde{u}({{\bf k},\omega})\,e^{i({\bf k}.{\bf x}-\omega t)}d^3k\,d\omega,\\
\nonumber
\zeta({\bf x},t)=\int\tilde{\zeta}({{\bf k},\omega})\,e^{i({\bf k}.{\bf x}-\omega t)}d^3k\,d\omega, \\
\nonumber
B_x({\bf x},t)=\int\tilde{B_x}({{\bf k},\omega})\,e^{i({\bf k}.{\bf x}-\omega t)}d^3k\,d\omega,\\
\nonumber
\zeta_B({\bf x},t)=\int\tilde{\zeta_B}({{\bf k},\omega})\,e^{i({\bf k}.{\bf x}-\omega t)}d^3k\,d\omega,\\
\eta_i({\bf x},t) = \int\tilde{\eta_i}({{\bf k},\omega})\,e^{i({\bf k}.{\bf x}-\omega t)}d^3k\,d\omega,
\label{four}
\end{eqnarray}
and substituting them into equations (\ref{orrv}), (\ref{zeta}), (\ref{orrb}) and (\ref{orrbzeta}) we obtain 
\begin{eqnarray}
\left(\begin{array}{cr}\tilde{u}({{\bf k},\omega})\\ 
\tilde{\zeta}({\bf k},\omega)\\\tilde{B}({{\bf k},\omega})\\\tilde{\zeta_B}({\bf k},\omega)\end{array}\right)={\cal M}^{-1}\left(\begin{array}{cr}\tilde{\eta_1}({\bf k},\omega)\\ 
\tilde{\eta_2}({\bf k},\omega)\\\tilde{\eta_3}({\bf k},\omega)\\\tilde{\eta_4}({\bf k},\omega)\end{array}\right),
\label{mat1}
\end{eqnarray}
where 
\begin{eqnarray}
{\cal M}=\left(\begin{array}{cr}{\cal M}_{11}\,\,\,\,\, {\cal M}_{12}\,\,\,\,\,{\cal M}_{13}\,\,\,\,\,{\cal M}_{14}\\ 
{\cal M}_{21}\,\,\,\,\, {\cal M}_{22}\,\,\,\,\,{\cal M}_{23}\,\,\,\,\,{\cal M}_{24}\\
{\cal M}_{31}\,\,\,\,\, {\cal M}_{32}\,\,\,\,\,{\cal M}_{33}\,\,\,\,\,{\cal M}_{34}\\
{\cal M}_{41}\,\,\,\,\, {\cal M}_{42}\,\,\,\,\,{\cal M}_{43}\,\,\,\,\,{\cal M}_{44}\end{array}\right),
\label{mat2}
\end{eqnarray}
\begin{eqnarray}
\nonumber
&&{\cal M}_{11}=ik^2\omega+ilk^2k_y-\frac{k^4}{R_e},\hspace{2mm}{\cal M}_{12}=\frac{2ik_z}{q},\hspace{2mm}{\cal M}_{13}=\frac{ik^2}{4\pi}(B_1k_y+k_z),\hspace{2mm}{\cal M}_{14}=0,\\
\nonumber
&&{\cal M}_{21}=ik_z\left(1-\frac{2}{q}\right),\hspace{2mm}{\cal M}_{22}=-i\omega-ilk_y+\frac{k^2}{R_e},\hspace{2mm}{\cal M}_{23}=0,\hspace{2mm}{\cal M}_{24}=\frac{-i}{4\pi}\left(B_1k_y+k_z\right),\\
\nonumber
&&{\cal M}_{31}=\left(-iB_1k_y-ik_z\right),\hspace{2mm}{\cal M}_{32}=0,\hspace{2mm}{\cal M}_{33}=\left(-i\omega-ilk_y+\frac{k^2}{R_m}\right),\hspace{2mm}{\cal M}_{34}=0,\\
&&{\cal M}_{41}=0,\hspace{2mm}{\cal M}_{42}=\left(-iB_1k_y-ik_z\right),\hspace{2mm}{\cal M}_{43}=-ik_z,\hspace{2mm}{\cal M}_{44}=\left(-i\omega-ilk_y+\frac{k^2}{R_m}\right),
\label{matrixcoeff}
\end{eqnarray}
where $\tilde{{\eta}_{i}}({\bf k},\omega)$ ($i=1,2,3,4$) are the components of noise in $k-\omega$ space ($k=\sqrt{k_x^2+k_y^2+k_z^2}$) with noise correlations given as follows

\begin{equation}
<\eta_i({\bf k},\omega) \eta_j({\bf k'},\omega')> = 2D_0 \delta^3({\bf k}+{\bf k'})\,\delta(\omega+\omega').
\label{noisecorr}
\end{equation}

\section{Autocorrelation functions in presence of non-zero noise cross-correlation}

In this section, we would analyze the spatiotemporal autocorrelations 
of the perturbation flow fields $u$, $\zeta$, $B_x$ and $\zeta_B$ for very large $R_e$ and 
$R_m$ in presence of non-zero noise cross-correlation. As mentioned earlier, this would imply the consideration of cross-variable affectation of flows; in other words, how symmetry structure could change due to non-trivial magnetohydrodynamical fluctuations. The choice of individually large $R_e$ and $R_m$ but a finite $\frac{R_e}{R_m}$ is quite meaningful for accretion flows in that this implies the presence of a finite Prandtl number . 

\subsection{Temporal autocorrelations}

The quantities of interest here are the following 

\begin{subequations}
\begin{eqnarray}
<u({\bf x},t)\,u({\bf x},t+\tau)> &=& C_{uu}(\tau)=\int d^3k\,d\omega\,e^{-i\omega\tau}
<\tilde{u}({\bf k},\omega)\,\tilde{u}(-{\bf k},-\omega)>,\\
<\zeta({\bf x},t)\,\zeta({\bf x},t+\tau)>&=& C_{\zeta \zeta}(\tau)=\int d^3k\,d\omega\,e^{-i\omega\tau}
<\tilde{\zeta}({\bf k},\omega)\,\tilde{\zeta}({-\bf k},-\omega)>,\\
<B_x({\bf x},t)\,B_x({\bf x},t+\tau)>&=& C_{BB}(\tau)=\int d^3k\,d\omega\,e^{-i\omega\tau}
<\tilde{B_x}({\bf k},\omega)\,\tilde{B_x}(-{\bf k},-\omega)>,\\
<\zeta_B({\bf x},t)\,\zeta_B({\bf x},t+\tau)>&=& C_{\zeta_B \zeta_B}(\tau)
=\int d^3k\,d\omega\,e^{-i\omega\tau}<\tilde{\zeta_B}({\bf k},\omega)\,\tilde{\zeta_B}(-{\bf k},-\omega)>,
\label{tempautocorr}
\end{eqnarray}
\end{subequations}

\noindent
calculated over the projected hyper-surface for which $k_x=k_y=k_z=\frac{|k|}{\sqrt{3}}$, as in \cite{nath2013}. This corresponds to a special choice of initial perturbation.
As our principal interest is in probing the scaling regime at which non-zero noise cross-correlations contribute to create (or otherwise destroy at times) sudden surges in energy activity, this restriction would only alter the magnitude of 
the probability density function (PDF) at worst while retaining the qualitative structure, including scaling exponents, if any, unchanged. This also adds up to the noise incompressibility
constraint while still retaining the non-trivial nature of the cross-correlations unaffected.

In line with \cite{nath2013}, we now perform the $\omega$-integration of the integrands in equation (\ref{tempautocorr}) by computing 
the four second order poles of the kernel which are functions of $k$. 
The form of all the integrands in equation (\ref{tempautocorr}) is given by
\begin{eqnarray}
\nonumber
f(k,\omega)=\frac{p(k,\omega)}{[\omega-\omega_1(k)]^2[\omega-\omega_2(k)]^2[\omega-\omega_3(k)]^2[\omega-\omega_4(k)]^2},
\end{eqnarray}
which clearly reveals second order imaginary poles at $\omega_1, \omega_2, \omega_3$ and $\omega_4$. 
We choose the range of $k$ in such a way that the poles lie in the upper-half of the complex plane.
The residue theorem is then used to calculate the frequency contributions at the poles
in the wave vector range $k_0$ to $k_m$, where $k_0=\frac{2\pi}{L_{\text{max}}}$,
$k_m=\frac{2\pi}{L_{\text{min}}}$ with $L=L_{\rm max}-L_{\rm min}$ 
being the size of the chosen small section of the flow (\enquote{small gap approximation}) in the radial direction 
(arbitrarily chosen to be $2$ units throughout for our evaluations),
we obtain $C_{uu}(\tau)$, $C_{\zeta \zeta}(\tau)$, $C_{B B}(\tau)$ and $C_{\zeta_B \zeta_B}(\tau)$.

\begin{figure}[H]
 \centering
\includegraphics[scale=0.9]{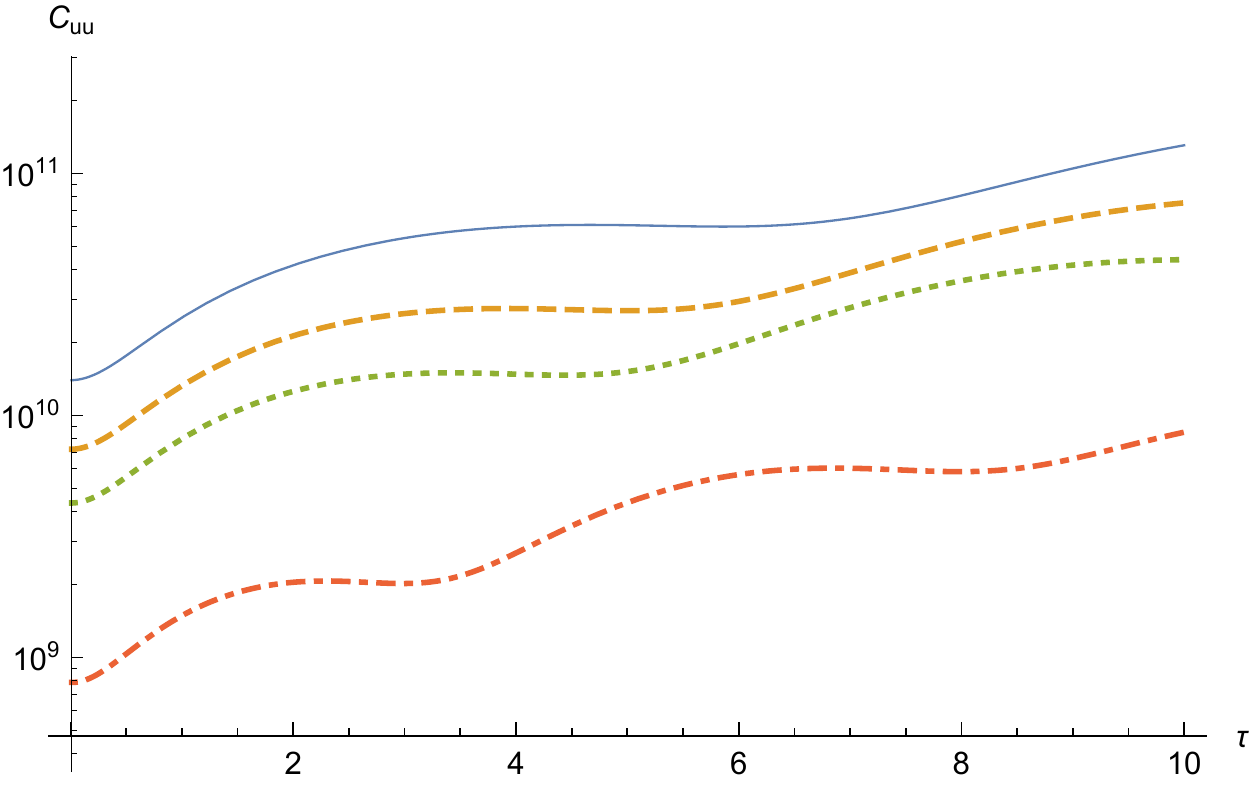}
\caption{(Color online) Temporal autocorrelations of velocity ($C_{uu}$) upto $\tau=10$. The solid line represents $q \approx 2$, the dashed line is for $q=1.7$, the dotted line represents $q=1.5$ and the dot-dashed line is for $q=1$. While oscillation amplitudes are lesser compared to the case without cross-correlations in noise \cite{nath2013}, each q-valued correlation spectrum comes up with a separate plot at varying magnitudes. This is remarkably different compared to the zero cross-correlation in noise situation where velocity-autocorrelation functions for all q-valued data collapsed over each other (Fig 3 in \cite{nath2013}).}
\label{uu_temp_10}
\end{figure}

\begin{figure}[H]
 \centering
\includegraphics[scale=0.9]{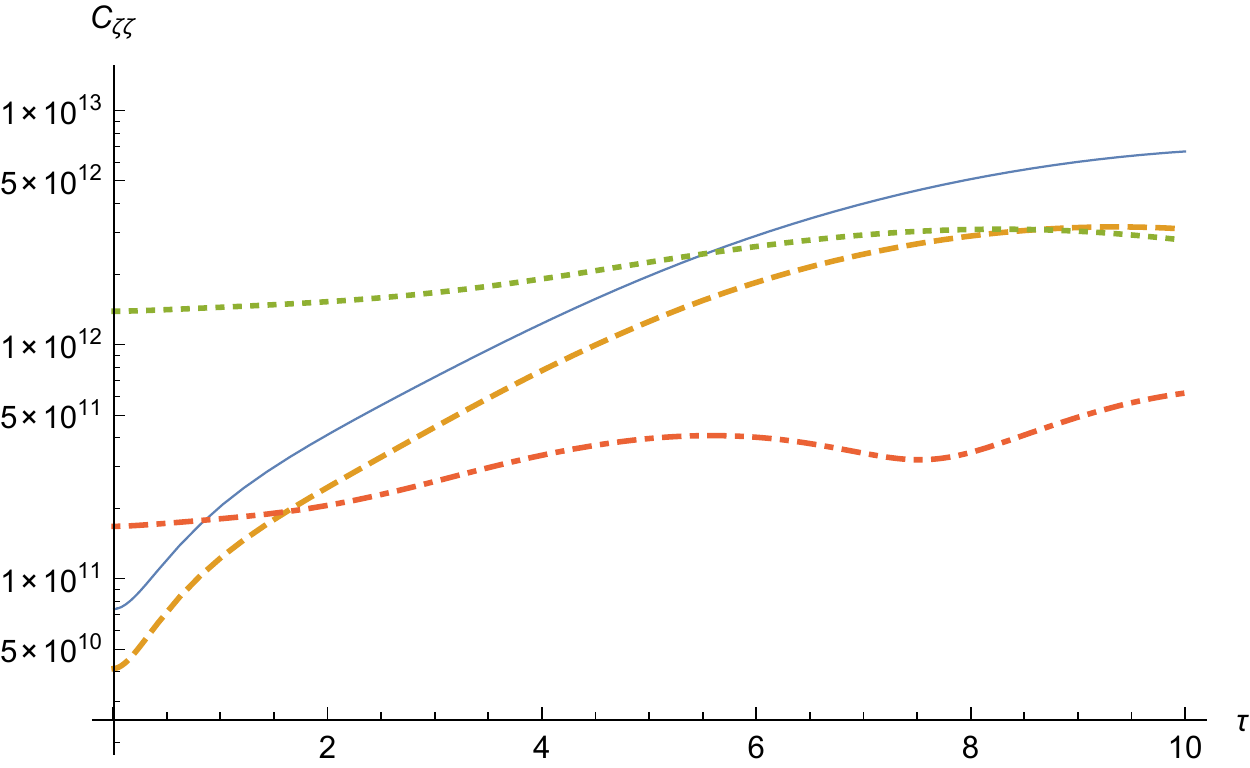}
\caption{(Color online) Temporal autocorrelations of vorticity ($C_{\zeta\zeta}$) upto $\tau=10$. Plot styles and symbol conventions are same as in Fig~\ref{uu_temp_10}. Once again the effects of noise cross-correlation can be clearly seen in that the plots for the different q-values do not collapse over each other. The other remarkable difference lies in the increasing profiles of vorticity correlations for the larger q-values ($q\approx 2$, $q=1.7$) while $q=1$ qualitatively matches with the profile with zero cross-correlation in noise. This indicates that neglecting noise cross-correlations implies a domination of the flat rotation profile over other available modes like Keplerian disc and constant angular momentum.}
\label{zz_temp_10}
\end{figure}

\begin{figure}[H]
 \centering
\includegraphics[scale=0.9]{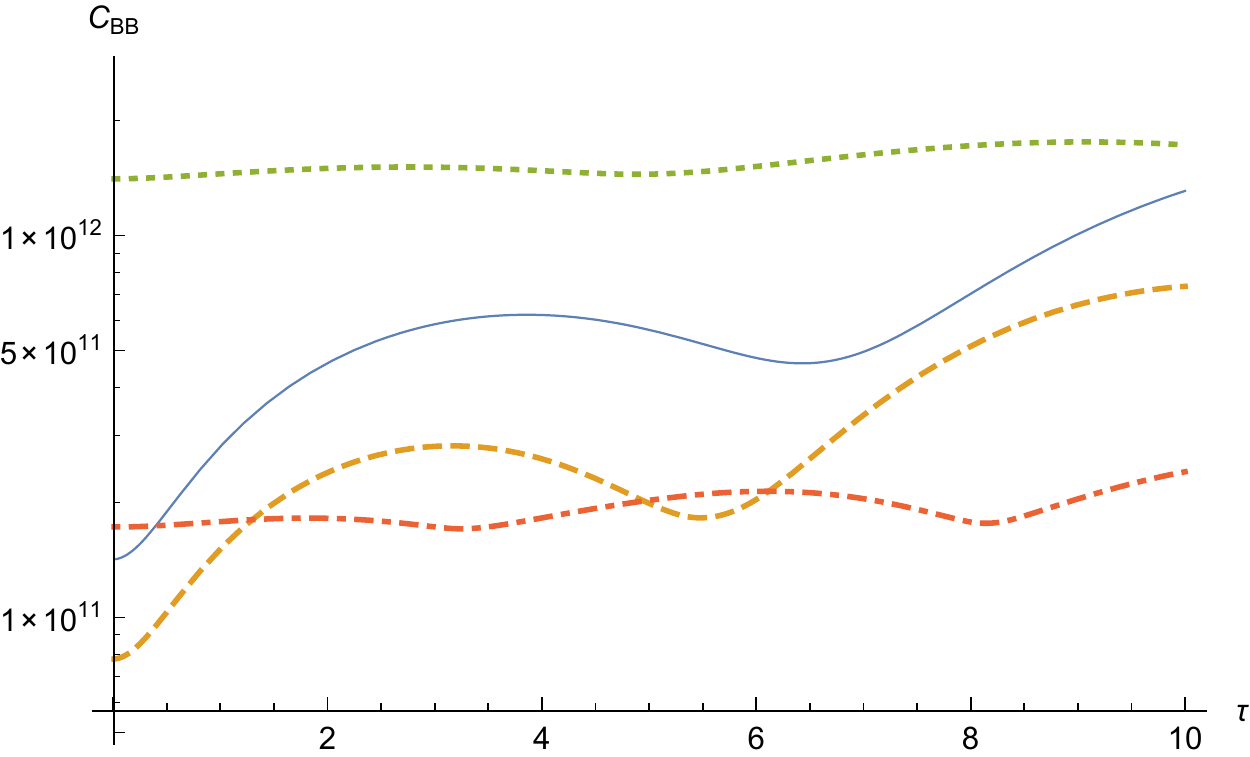}
\caption{(Color online) Temporal autocorrelations of the magnetic field ($C_{BB}$) upto $\tau=10$. Plot styles and symbol conventions are same as in Fig~\ref{uu_temp_10}. As opposed to the two other autocorrelation plots in presence of non-zero cross-correlation in noise, as shown above, the order of magnitude remains the same although higher q-values ($q\approx 2$, $q=1.7$) decidedly show faster and stronger time oscillating profiles (Alfven waves) unlike in the previous (zero noise-correlation) case.}
\label{bb_temp_10}
\end{figure}

\begin{figure}[H]
 \centering
\includegraphics[scale=0.9]{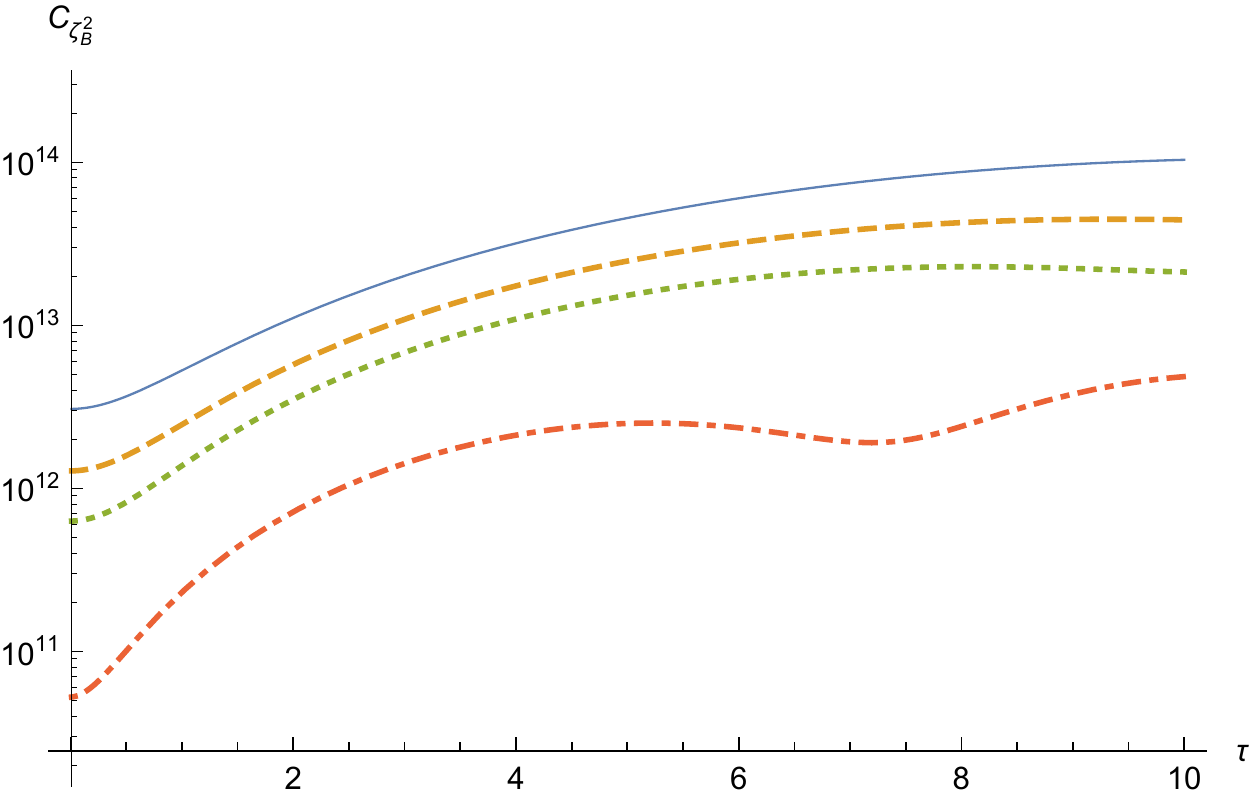}
\caption{(Color online) Temporal autocorrelations of magnetic vorticity ($C_{\zeta_B^2}$) upto $\tau=10$. Plot styles and symbol conventions are same as in Fig~\ref{uu_temp_10}. The oscillation profiles clearly indicate similarity with the velocity autocorrelation function shown in Fig \ref{uu_temp_10} above. While oscillations are much lesser compared to the case without cross-correlations in noise, each q-value comes up with a separate plot with periodic oscillations appearing for the flat rotation curve (q=1). This is remarkably different compared to the zero cross-correlation in noise situation where velocity-autocorrelation functions for all q-valued data collapsed over each other.}
\label{zbzb_temp_10}
\end{figure}

The two-point temporal autocorrelation functions that we get individually for velocity-velocity (Fig~\ref{uu_temp_10}), vorticity-vorticity (Fig~\ref{zz_temp_10}), magnetic field-magnetic field (Fig~\ref{bb_temp_10}) and magnetic vorticity-magnetic vorticity (Fig~\ref{zbzb_temp_10}) are most suggestive, especially when compared to similar results but in absence of any noise cross-correlation (detailed in \cite{nath2013}). 

\noindent
As discussed in details in \cite{nath2013}, all values of $q$ that are sufficiently close to the $q\to 2$ limit, irrespective of the exact value of choice being $q=1.9999$ or $q=1.99999$, etc. the corresponding correlation plots simply superpose on top of each other. The more interested readers could avail the detailed analysis of Figures 3, 9 and 17 in this reference \cite{nath2013}. All results presented in this article are for $q=1.9999$ but the plots remain both qualitatively and quantitatively unchanged for even larger values of $q$ closer to the limit $q=2$. In order to refrain from repetition, we avoid 

\noindent
A remarkable impact of non-zero noise cross-correlation in the spatiotemporal dynamics is in the development of low frequency oscillatory waves, called Alfven waves \cite{ll05,Alfven}, due to interaction of the magnetic flux with the extra inertia generated by the noise coupling, essentially leading to (oscillatory) instability due to additional energy influx. The resultant renormalized phase velocity pumps even more energy in to the system that adds up to the quantitative influx. This is the reason why even though no new phenomenological changes are observed for the case of the spatial autocorrelation functions (next section), unlike that for temporal dynamics, the energy fluxes are hugely revitalized resulting in orders of magnitude difference (spatial change in the energy flux is proportional to the the spatial two-point autocorrelation function, as was previously shown in \cite{chattopadhyay2013}) in energy flows.

In Fig~\ref{uu_temp_10}, apart from the flat rotation curve ($q=1$), all higher q-valued correlation profiles show increasingly smaller amplitudes and larger wave lengths for the resultant Alfven waves at low k-values. Apart from an overall difference of up to two orders of magnitude for higher q-valued correlation spectra (higher with non-zero noise correlation), the small k-profiles ($C_{uu}(k\to0)$) show a steep gradient increasing with the value of q. The only exception is with $q=1$ where due to velocity conservation (instead of angular momentum conservation or Keplerian rotation), the qualitative profile remains relatively unchanged with the only change showing up in the number of oscillating cycles (that is number of Alfven waves) and their wave lengths which are now stretched due to non-zero noise cross-correlation.

\noindent
For all values of $q\neq 1$, the vorticity-vorticity autocorrelation spectra (Fig~\ref{zz_temp_10}) assume remarkably different qualitative forms compared to the case without any noise cross-correlation. What the noise here does is to initially aggravate the Coriolis flux resulting in increasing gradients for each of the correlations functions ($C_{\zeta \zeta}$) for $q\neq1$ that then saturate at slightly higher magnitudes, at which point the zero noise-correlation dynamics takes over. Once again, $q=1$ turns out to be a special case in that it practically replicates the zero noise correlation structure. This is not difficult to perceive and could be attributed to an effective lack of momentum conservation whose structure changes with increasing angular momentum flux due to increased Coriolis forces.

\noindent
The magnetic field and magnetic vorticity autocorrelation functions too show distinctly different qualitative and quantitative characteristics (compared to the zero noise cross-correlation case as in \cite{nath2013}). Evidently more energy coming from noise cross-correlations induce stronger Alfven waves as are evident from comparatively fast oscillating profiles in the magnetic field correlation function (Fig~\ref{bb_temp_10}). The magnetic vorticity autocorrelation though does not show much qualitative change compared to the zero noise cross-correlation case since here the fixation of the Prandtl number contrives to balance this extra rotational energy being pitched in to the dynamics (Fig~\ref{zbzb_temp_10}).

\subsection{Spatial autocorrelations}

The quantities of interest here are the following 

\begin{subequations}
\begin{eqnarray}
<u({\bf x},t)\,u({\bf x+r},t)> &=& S_{uu}(r)=\int d^3k\,d\omega\,e^{i{\bf k}.{\bf r}}
<\tilde{u}({\bf k},\omega)\,\tilde{u}(-{\bf k},-\omega)>,\\
<\zeta({\bf x},t)\,\zeta({\bf x+r},t)>&=& S_{\zeta \zeta}(r)=\int d^3k\,d\omega\,e^{i{\bf k}.{\bf r}}
<\tilde{\zeta}({\bf k},\omega)\,\tilde{\zeta}({-\bf k},-\omega)>,\\
<B_x({\bf x},t)\,B_x({\bf x+r},t)>&=& S_{BB}(r)=\int d^3k\,d\omega\,e^{i{\bf k}.{\bf r}}
<\tilde{B_x}({\bf k},\omega)\,\tilde{B_x}(-{\bf k},-\omega)>,\\
<\zeta_B({\bf x},t)\,\zeta_B({\bf x+r},t)>&=& S_{\zeta_B \zeta_B}(r)
=\int d^3k\,d\omega\,e^{i{\bf k}.{\bf r}}<\tilde{\zeta_B}({\bf k},\omega)\,\tilde{\zeta_B}(-{\bf k},-\omega)>.
\label{spatialautocorr}
\end{eqnarray}
\end{subequations}

\noindent
The three dimensional spatial integration may be reduced to the radial component only. Following is an example:

\begin{equation}
S_{uu}(r) 
= 2\pi\int_{k_0}^{k_m}~dk~k^2 ~\int_0^\pi~d\theta~e^{ikr\cos\theta}~
\int d\omega~<\tilde{u}({\bf k},\omega)\,
\tilde{u}({-\bf k},-\omega)>,
\label{spatialautopcorr}
\end{equation}

The other autocorrelations would follow suit.

\begin{figure}[H]
 \centering
\includegraphics[scale=0.9]{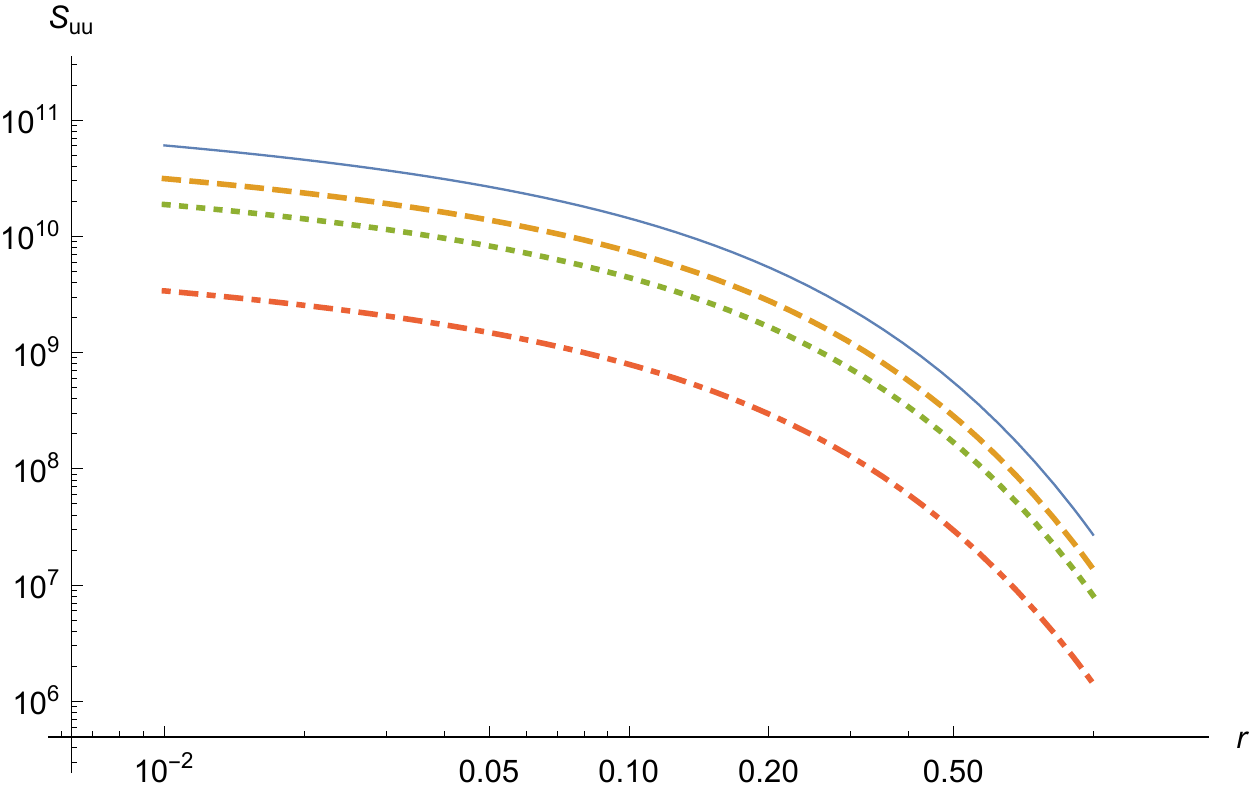}
\caption{(Color online) Spatial autocorrelations of velocity ($S_{uu}$) upto $r=1.0$. Plot styles and symbol conventions are same as in Fig~\ref{uu_temp_10}.}
\label{uu_spatial}
\end{figure}

\noindent
Apart from noting the obvious similarity with the zero noise cross-correlation case, introduction of noise cross-correlation does not seem to change much in the dynamics qualitatively. The qualitative similarity is not so very difficult to predict given the fact that a white Gaussian noise is not expected to renormalize the spatial autocorrelation spectrum. 

\begin{figure}[H]
 \centering
\includegraphics[scale=0.9]{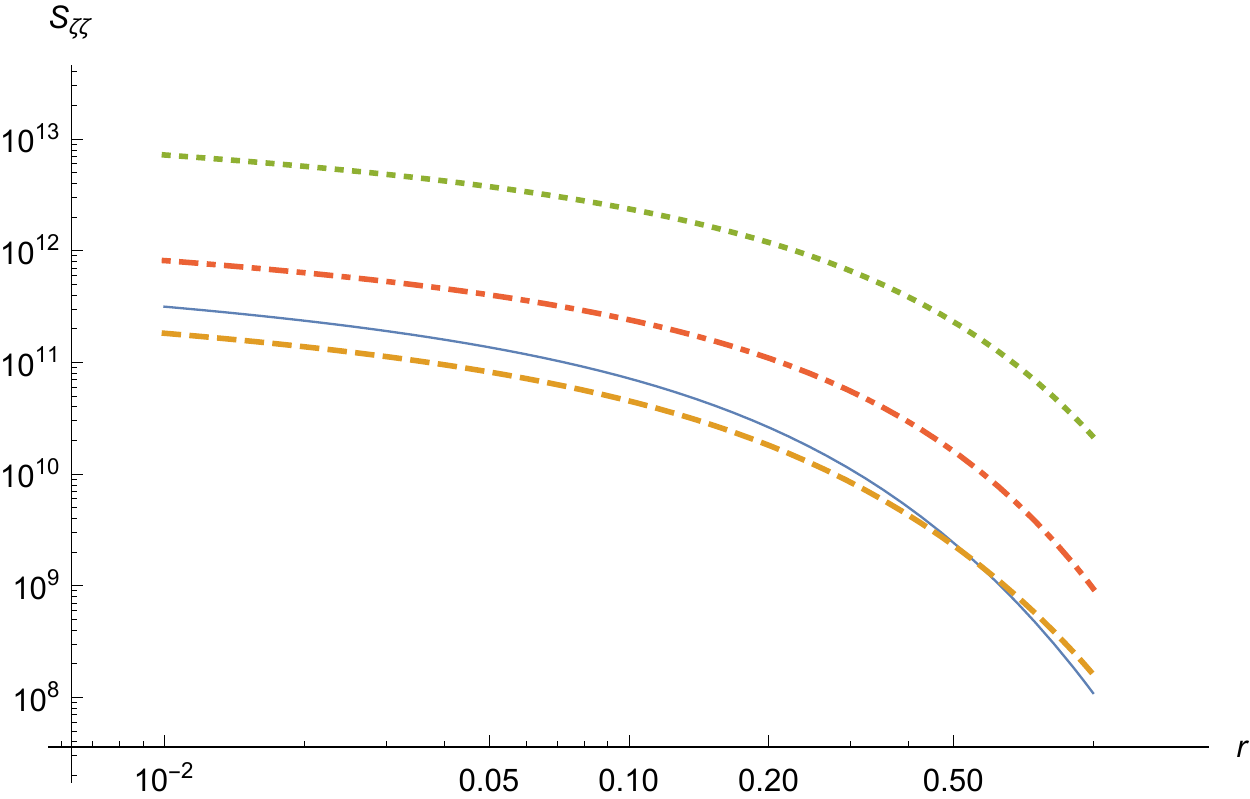}
\caption{(Color online) Spatial autocorrelations of vorticity ($S_{\zeta\zeta}$) upto $r=1.0$. Plot styles and symbol conventions are same as in Fig~\ref{uu_temp_10}.}
\label{zz_spatial}
\end{figure}

\noindent
The quantitative comparison is more interesting though. While the magnitude of energy spurt (${10}^{10}$ to ${10}^{11}$) remains roughly unchanged, what the noise cross-correlation now does is to define the spatial universality class while also clearly indicating that the transition from small-r to large-r spectrum (resulting from radial movement across the sheared surface) leads only to a \enquote{crossover} and is not a phase transition, a question that was not entirely settled in absence of noise cross-correlations \cite{nath2013}. Two additional spatial autocorrelation functions for non-trivial cross-correlated noise have been added in the supplemental material (Figures 1 and 2) \cite{supplemental}. The first of these figures show the spatial autocorrelation for magnetic vorticity while the second shows the same for magnetic field, each for a set of q-values as detailed in the legend to these figures.

The other similarity to be noted in between the two cases (noise cross-correlated versus zero cross-correlation in noise) is the fact that the autocorrelation functions in the log-log scale show almost parallel lines indicating identical gradients on both sides of the crossover in both cases. 

\section{Cross-correlation functions in presence of non-zero noise cross-correlation}

In this section, we would analyze the spatiotemporal cross-correlations 
of the perturbation flow fields $u$, $\zeta$, $B_x$ and $\zeta_B$ for very large $R_e$ and 
$R_m$ in presence of non-zero noise cross-correlation. The essential implication of such an estimation is to ascertain how much of a fluctuation in any one variable affects a different variable. Contrary to the drastic assumption of dropping all noise cross-correlations as in \cite{nath2013}, now we would consider contributions from such symmetry violating terms as well. As shown in the context of a sheared boundary layer flow \cite{akc_shear}, such quantities may dramatically alter both qualitative and quantitative outcomes in variable cross-correlations. Using \cite{nath2013} as our benchmark (with zero noise cross-correlation), the results analyzed in this section would be contrasted against the report in \cite{nath2013}. 

\subsection{Temporal cross-correlations}

The quantities of interest here are the following 

\begin{subequations}
\begin{eqnarray}
<u({\bf x},t)\,B_x({\bf x},t+\tau)> &=& C_{uB}(\tau)=\int d^3k\,d\omega\,e^{-i\omega\tau}
<\tilde{u}({\bf k},\omega)\,\tilde{B_x}(-{\bf k},-\omega)>\\
<u({\bf x},t)\,\zeta({\bf x},t+\tau)>&=& C_{u \zeta}(\tau)=\int d^3k\,d\omega\,e^{-i\omega\tau}
<\tilde{u}({\bf k},\omega)\,\tilde{\zeta}({-\bf k},-\omega)>\\
<u({\bf x},t)\,\zeta_B({\bf x},t+\tau)>&=& C_{u \zeta_B}(\tau)=\int d^3k\,d\omega\,e^{-i\omega\tau}
<\tilde{u}({\bf k},\omega)\,\tilde{\zeta_B}(-{\bf k},-\omega)>\\
<\zeta({\bf x},t)\,\zeta_B({\bf x},t+\tau)>&=& C_{\zeta \zeta_B}(\tau)
=\int d^3k\,d\omega\,e^{-i\omega\tau}<\tilde{\zeta}({\bf k},\omega)\,\tilde{\zeta_B}(-{\bf k},-\omega)>\\
<B_x({\bf x},t)\,\zeta_B({\bf x},t+\tau)>&=& C_{B \zeta_B}(\tau)
=\int d^3k\,d\omega\,e^{-i\omega\tau}<\tilde{B_x}({\bf k},\omega)\,\tilde{\zeta_B}(-{\bf k},-\omega)>\\
<\zeta({\bf x},t)\,B_x({\bf x},t+\tau)>&=& C_{B \zeta}(\tau)
=\int d^3k\,d\omega\,e^{-i\omega\tau}<\tilde{\zeta}({\bf k},\omega)\,\tilde{B_x}(-{\bf k},-\omega)>.
\label{tempcrosscorr}
\end{eqnarray}
\end{subequations}

\noindent
The magnitude of the two-point temporal cross-correlation function is higher than before and oscillation is less compared to the zero noise cross-correlation case. For cross-correlated variables, the extra energy generated due to noise cross-correlation often leads to effective boost in energy in one of the variables while effectively suppressing the dynamics of the other through slower growth, resulting in a dynamical \enquote{push-pull} mechanism. 

\begin{figure}[H]
 \centering
\includegraphics[scale=0.9]{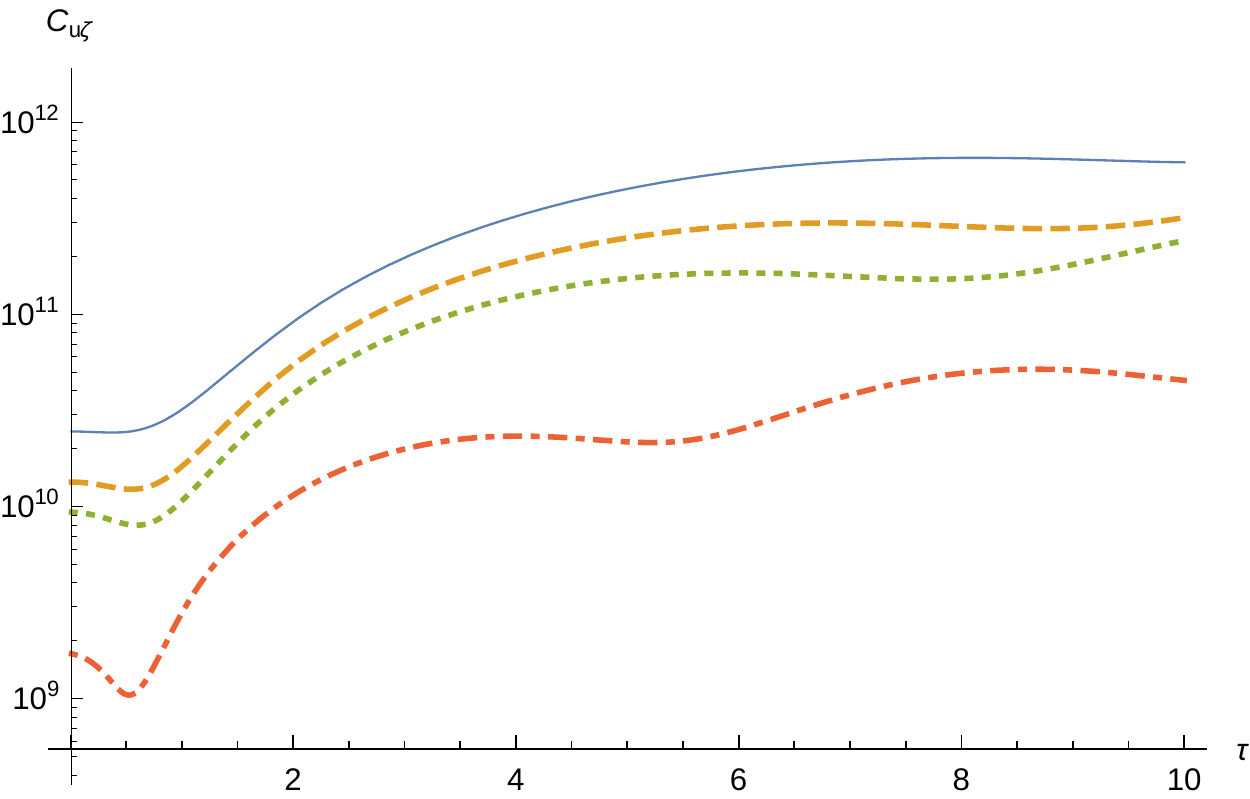}
\caption{(Color online) Temporal cross-correlation of velocity and vorticity ($C_{u\zeta}$). Plot styles and symbol conventions are same as in Fig~\ref{uu_temp_10}.}
\label{uz_temp_10}
\end{figure}

\noindent
In the case of the velocity-vorticity cross-correlation, it is obvious that the comparative boost in vorticity energy is higher than that for velocity. As a result of this \enquote{competition}, after an initially time growing profile, the (velocity-vorticity) cross-correlation function shows a clear saturation unlike the case without any noise cross-correlation. An indirect confirmation of this lies in the magnitude of the saturation regime that shows an order of magnitude lower value compared to the case without any noise cross-correlation. For the same reason, the Alfven oscillations for large time scales are suppressed here. For small enough time scales, up to a specific cut-off, the rate of energy dissipation for the velocity part outperforms the energy growth rate due to vorticity. This is the reason for the initial dip in the cross-correlation profiles (for all values of $q$). Beyond a critical time scale though, the vorticity factor starts dominating the energy picture which then accounts for the follow-up growth.

\begin{figure}[H]
 \centering
\includegraphics[scale=0.9]{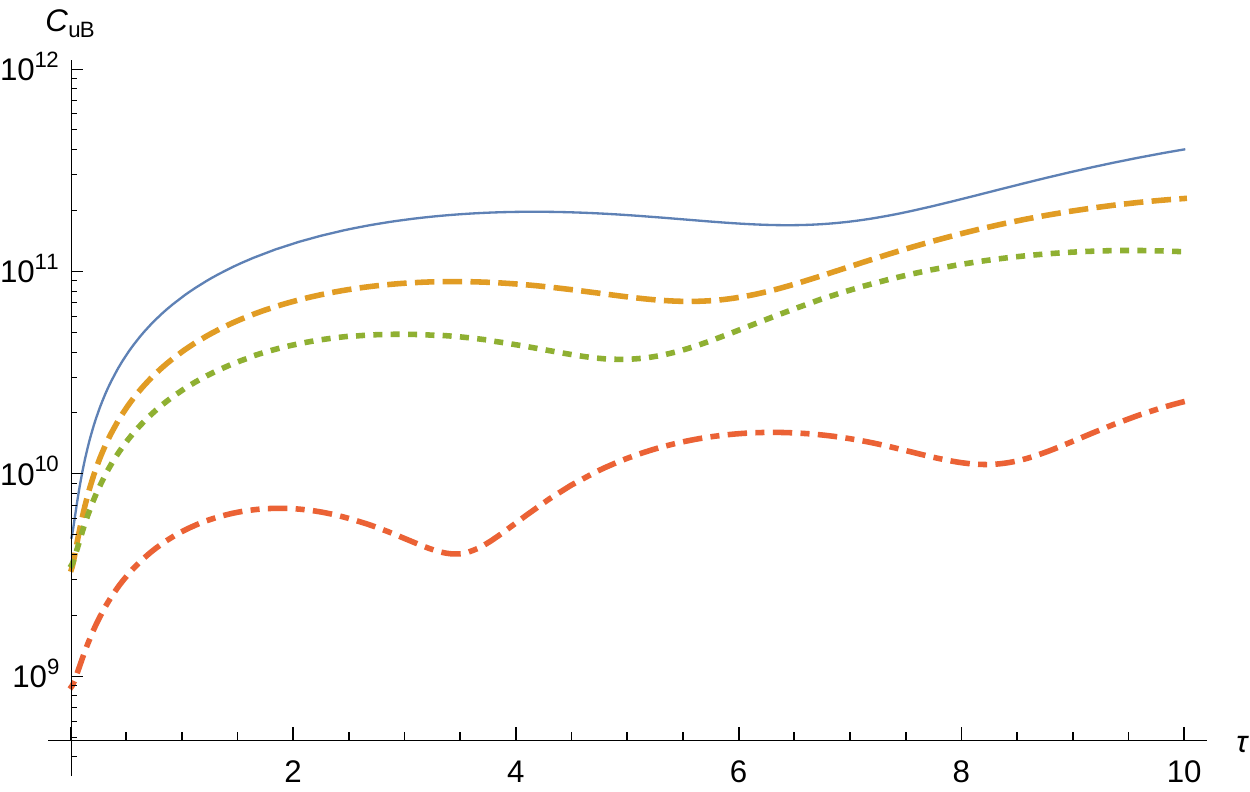}
\caption{(Color online) Temporal cross-correlation of velocity and magnetic field ($C_{uB}$). Magnitude is higher than the zero noise cross-correlation counterpart and (Alfven \cite{Alfven}) oscillation is also there. Extra cross-correlation in noise is equivalent to addition of extra energy in the system in presence of velocity-vorticity cross-correlation. This extra energy downplays the previous energy by destroying the eddies previously generated. This results in lower oscillations. Plot styles and symbol conventions are same as in Fig~\ref{uu_temp_10}.}
\label{ub_temp_10}
\end{figure}

While being significantly different to their equivalent zero noise cross-correlation cases, the rest of the cross-correlation functions mostly demonstrate traditional profiles for temporal correlation functions with a sharp increase in value for small times followed by respective saturation at larger time scales. In line with the explanation provided earlier, the difference in the comparative relaxation time scales of the cross-correlating variables lead to competitions in their mutual rates of growth, thereby ensuring lower saturation regimes for each noise cross-correlated case compared to their noise-cross-correlation-free equivalents. In all these plots the $q=1$ structure provides the strongest signature of oscillating Alfven waves, that once again agrees with the extra energy input due to nonzero noise cross-correlation hypothesis propounded earlier.

\begin{figure}[H]
 \centering
\includegraphics[scale=0.9]{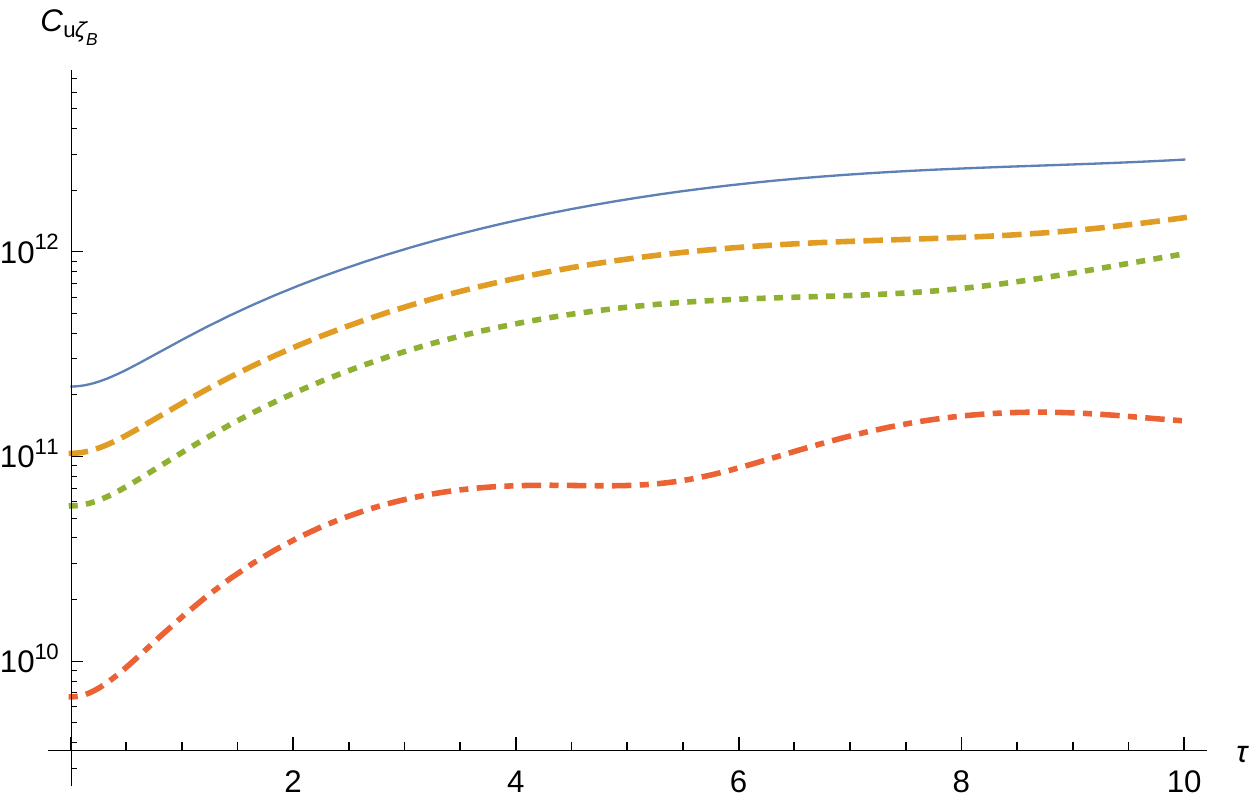}
\caption{(Color online) Temporal cross-correlation of velocity and magnetic vorticity ($C_{u\zeta_B}$). Magnitude is almost at the same range as in the zero noise cross-correlation counterpart but almost without any oscillation. Plot styles and symbol conventions are same as in Fig~\ref{uu_temp_10}.}
\label{uzb_temp_10}
\end{figure}

\begin{figure}[H]
 \centering
\includegraphics[scale=0.9]{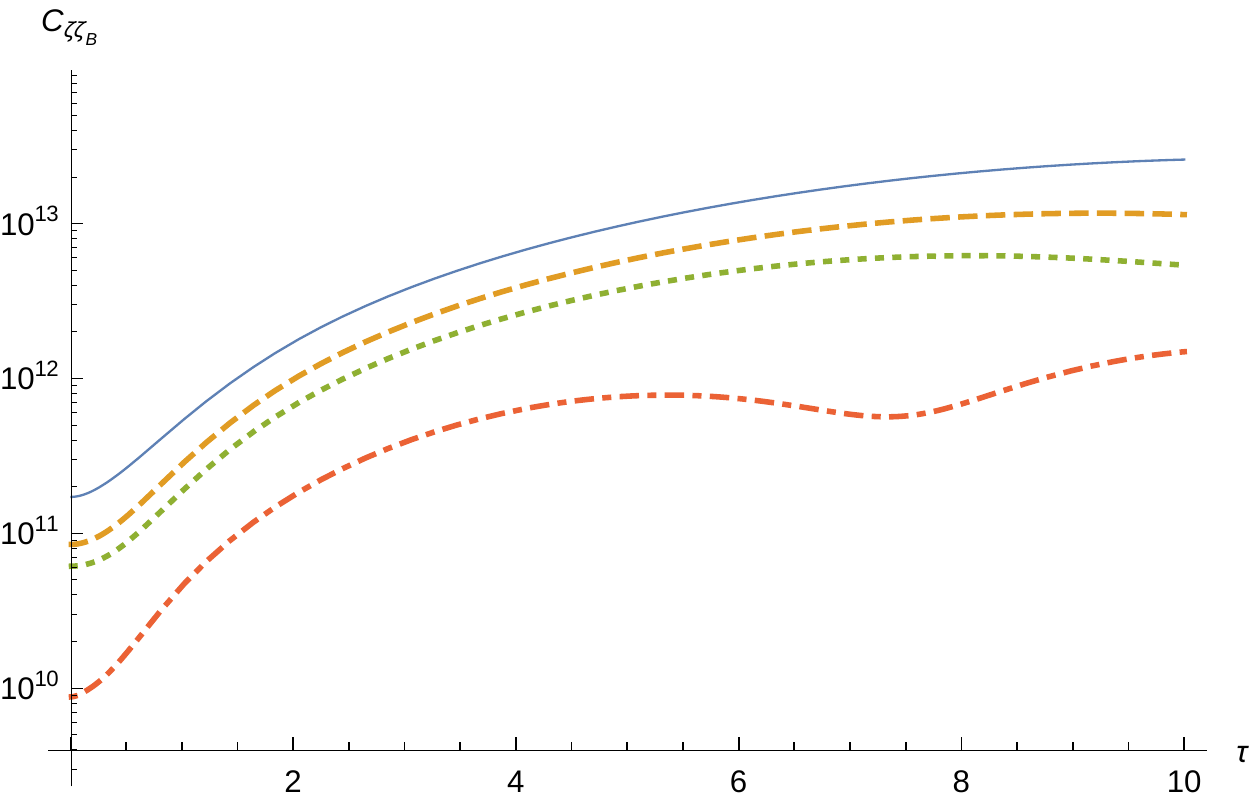}
\caption{(Color online) Temporal cross-correlation of vorticity and magnetic vorticity ($C_{\zeta\zeta_B}$). Magnitude is higher than zero noise cross-correlation counterpart and oscillation is there as previous. Plot styles and symbol conventions are same as in Fig~\ref{uu_temp_10}.}
\label{zzb_temp_10}
\end{figure}

\begin{figure}[H]
 \centering
\includegraphics[scale=0.9]{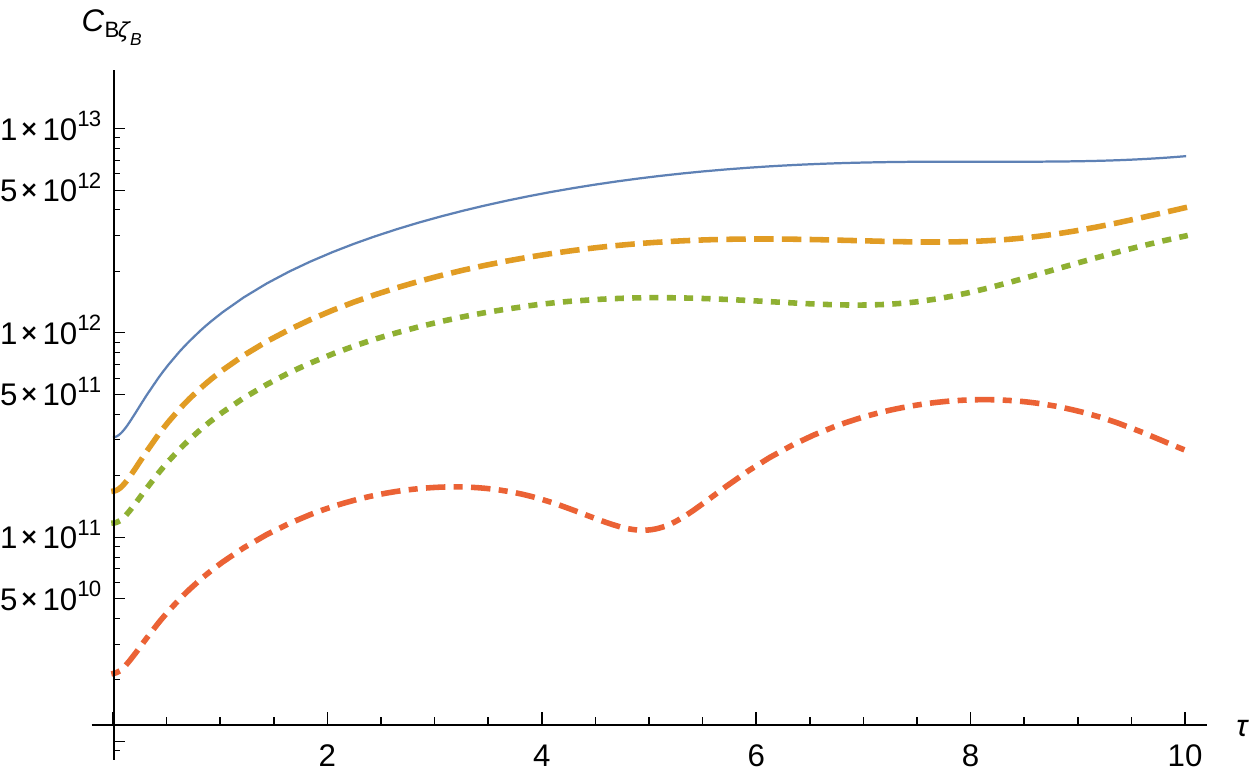}
\caption{(Color online) Temporal cross-correlation of magnetic field and magnetic vorticity ($C_{B\zeta_B}$). Magnitude is almost same (may be little higher but not significant) and (Alfven \cite{Alfven}) oscillation is also not that much different than zero noise cross-correlation counterpart. Plot styles and symbol conventions are same as in Fig~\ref{uu_temp_10}.}
\label{bzb_temp_10}
\end{figure}

\begin{figure}[H]
 \centering
\includegraphics[scale=0.9]{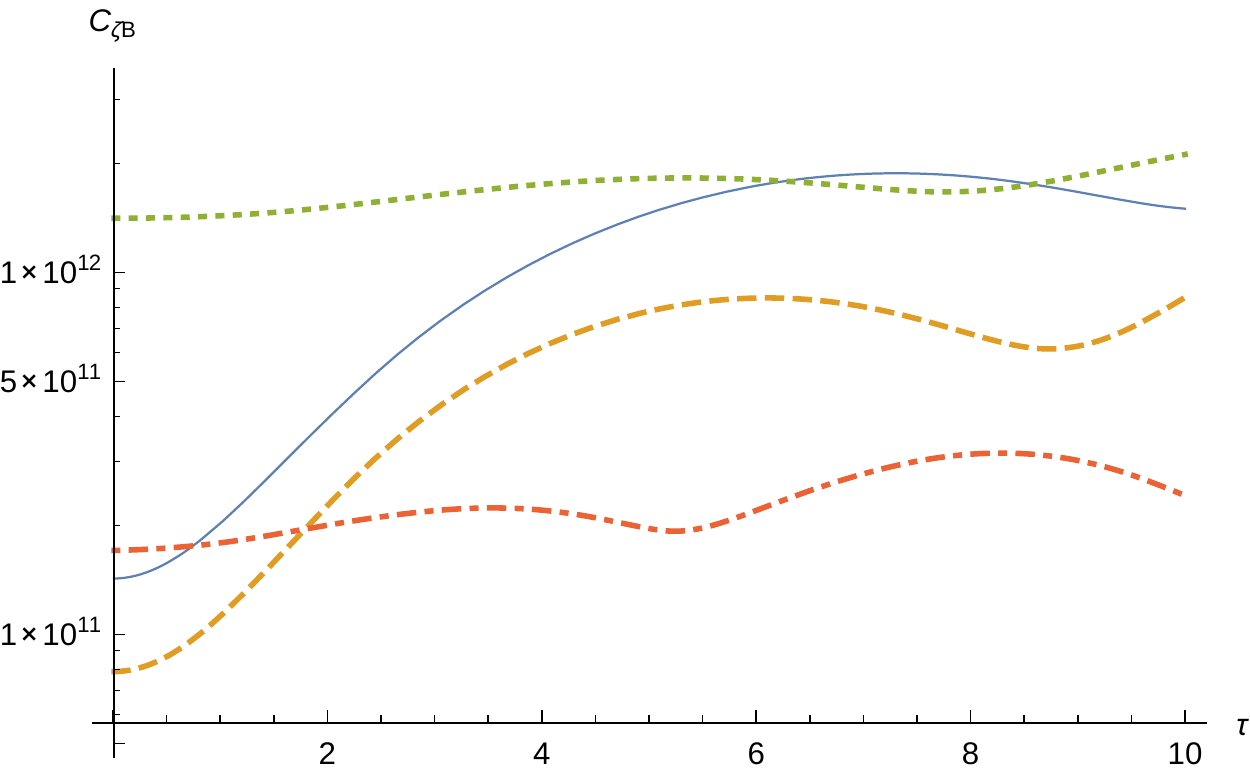}
\caption{(Color online) Temporal cross-correlation of vorticity and magnetic field ($C_{\zeta B}$). Plot styles and symbol conventions are same as in Fig~\ref{uu_temp_10}.}
\label{magvor_temp_10}
\end{figure}

\noindent
In order to address the issue of increasing proximity to the $q \to 2$ limit, we have added a figure in the (online) supplemental section \cite{supplemental} that shows the temporal cross correlation between velocity and magnetic field for 6 different values of $q$: $q=1.9999$, $q=1.99999$, $q=1.999999$, $q=1.7$, $q=1.5$ and $q=1.0$. This figure convincingly clarifies that apart from linear shifts in the values of the cross-correlation function, the $q \to2$ limit remains qualitatively unchanged for all q values nearing 2. As a sider, it may be added, that the same feature could be associated with all other cross and auto correlation functions that we do not show to avoid repetitions. 

\subsection{Spatial cross-correlations}

The quantities of interest here are the following 

\begin{subequations}
\begin{eqnarray}
<u({\bf x},t)\,B_x({\bf x+r},t)> &=& S_{uB}(r)=\int d^3k\,d\omega\,e^{i{\bf k}.{\bf r}}
<\tilde{u}({\bf k},\omega)\,\tilde{B_x}(-{\bf k},-\omega)>\\
<u({\bf x},t)\,\zeta({\bf x+r},t)>&=& S_{u \zeta}(r)=\int d^3k\,d\omega\,e^{i\omega\tau}
<\tilde{u}({\bf k},\omega)\,\tilde{\zeta}({-\bf k},-\omega)>\\
<u({\bf x},t)\,\zeta_B({\bf x+r},t)>&=& S_{u \zeta_B}(r)=\int d^3k\,d\omega\,e^{i\omega\tau}
<\tilde{u}({\bf k},\omega)\,\tilde{\zeta_B}(-{\bf k},-\omega)>\\
<\zeta({\bf x},t)\,\zeta_B({\bf x+r},t)>&=& S_{\zeta \zeta_B}(r)
=\int d^3k\,d\omega\,e^{i\omega\tau}<\tilde{\zeta}({\bf k},\omega)\,\tilde{\zeta_B}(-{\bf k},-\omega)>\\
<B_x({\bf x},t)\,\zeta_B({\bf x+r},t)>&=& S_{B \zeta_B}(r)
=\int d^3k\,d\omega\,e^{i\omega\tau}<\tilde{B_x}({\bf k},\omega)\,\tilde{\zeta_B}(-{\bf k},-\omega)>\\
<\zeta({\bf x},t)\,B_x({\bf x+r},t)>&=& S_{B \zeta}(r)
=\int d^3k\,d\omega\,e^{i\omega\tau}<\tilde{\zeta}({\bf k},\omega)\,\tilde{B_x}(-{\bf k},-\omega)>.
\label{spatialcrosscorr}
\end{eqnarray}
\end{subequations}

\noindent
As before, the three dimensional spatial integration may be reduced to the radial component only. Following is an example (one of the six possible):

\begin{equation}
S_{uB}(r) 
= 2\pi\int_{k_0}^{k_m}~dk~k^2 ~\int_0^\pi~d\theta~e^{ikr\cos\theta}~
\int d\omega~<\tilde{u}({\bf k},\omega)\,
\tilde{B_x}({-\bf k},-\omega)>,
\label{spatialcorr}
\end{equation}

\noindent
As like the case with the spatial autocorrelation functions, the spatial cross-correlations too do not show any significant qualitative difference compared to their zero noise-correlated counterparts. While in a way this makes the spatial profiles \enquote{uninteresting}, in the sense of being less exciting, they immensely help in retaining the sanctity of the physical logic presented earlier.

\begin{figure}[H]
 \centering
\includegraphics[scale=0.9]{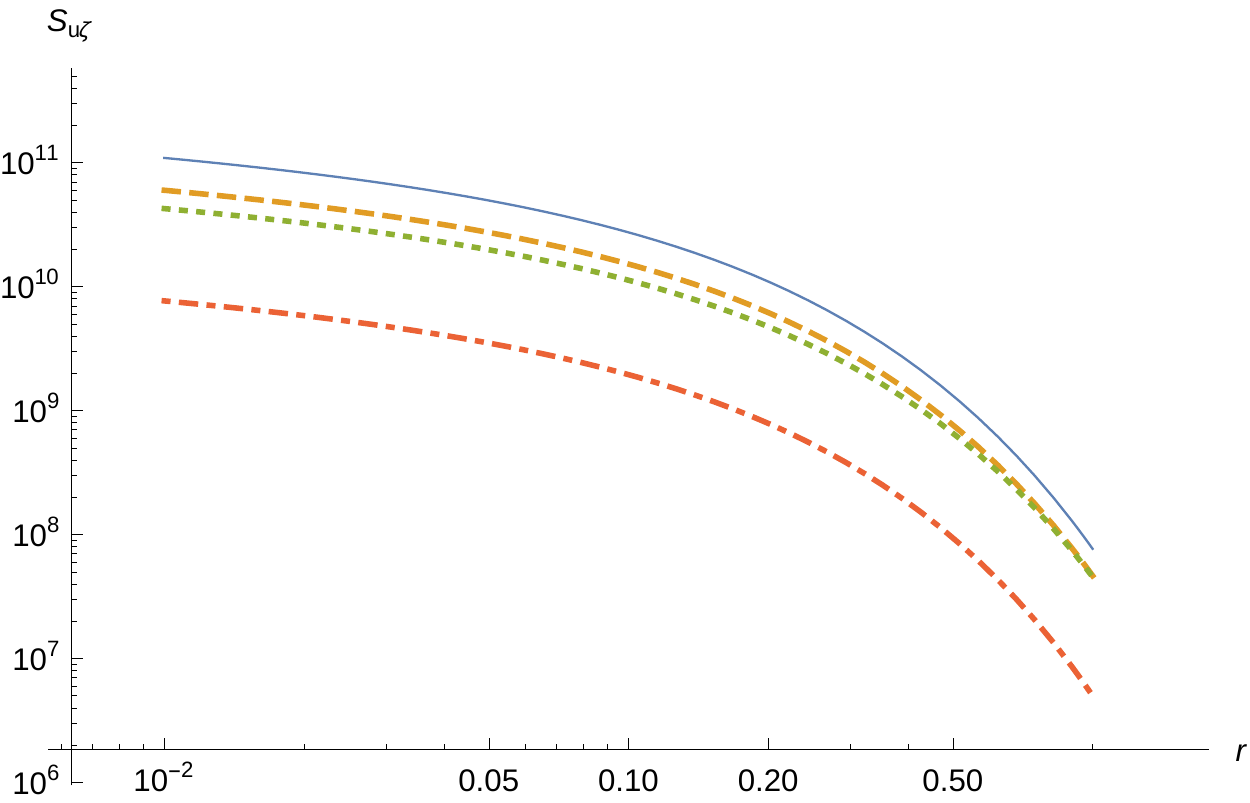}
\caption{(Color online)Spatial cross-correlation of velocity and vorticity ($S_{u\zeta}$) upto $\tau=10$. Huge difference in magnitude with zero cross-correlated noise counterpart \cite{nath2013}. At least 9 orders of magnitude difference. Plot styles and symbol conventions are same as in Fig~\ref{uu_temp_10}.}
\label{uz_spatial}
\end{figure}

\noindent
Only one spatial correlation plot is presented in the main text to indicate the quantitative variation brought about by the noise cross correlation in spatial dynamics. Other spatial plots while not being qualitatively any different than previous work \cite{nath2013}, they still do indicate varying levels of quantitative difference of previous work \cite{nath2013} as functions of system parameters ($R_e, R_m, q$, etc.). Five additional spatial cross-correlation functions for non-trivial cross-correlated noise have been added in the supplemental material (Figures 3 to 7) \cite{supplemental}. Once again, they do not indicate any qualitative or behavioral change, rather all changes are limited to quantitative variations.

A non-zero value for noise cross-correlation implies a violation of time translation symmetry making it imperative that only temporal correlation functions, both autocorrelation and cross-correlations, should be the primarily affected ones. This is what we have seen already. In order to enact similar qualitative changes in spatial correlation profiles, we needed to have a spatially correlated colored noise that would have ensured a space reflection symmetry violation. This is outside of our present remit as well as interest too.

\section{Summary and conclusions}

In this paper, our primary focus has been the extension of the groundwork laid down in our previous publications \cite{chattopadhyay2013,nath2013} where we established an alternative theoretical hypothesis (compared to \cite{balbus_hawley1996,balbus_hawley1999,man2005}) that is capable of explaining the origin of instability in the Rayleigh stable limit often leading to turbulence both in magnetized, rotating and non-rotating shear flows supplemented by Coriolis 
force), now also incorporating the effects of nonzero noise cross-correlation across all four variables (velocity, vorticity, magnetic field and magnetic vorticity). The target application is in the transport properties of accretion flows, both cold and hot. 

The importance of thermal noise as a (field theoretically) \enquote{relevant} order parameter was already established in these previous works (as also in \cite{paoletti,matsakos2013,proga2007}, the latter focusing on general thermal perturbations and is not expressly stochastic though) but what was not addressed in either publication is the role of time symmetry violation (both translational and reflectional) in the energy growth dynamics, that could potentially obviate or otherwise nullify the impact of the noise induced dynamics through (constructive or destructive, as the case might be) interference of additional energy flux introduced into the system by time symmetry violating nonzero noise cross-correlations. While we have no claim towards a unique mechanism of studying subcritical transition to turbulence in time symmetry violating Keplerian accretion disks \cite{rincon2007,maretzke2014}, the mechanism introduced here is consistent with our approach established earlier \cite{chattopadhyay2013,nath2013} in effecting such symmetry violations through stochastic noise distributions. 

By introducing noise cross-correlation as a time symmetry violating measure, what we have essentially done is to create additional influx or efflux of energy into the magnetized rotating accretion flows such that at any transient time scale, there is either a drop or increase in the total energy of the system on top of sheared dissipation. While such a cross-correlation in noise boosts the effective Coriolis force expressing itself as an added energy flux, there is an effective deconstruction with respect to the linear velocity flow component. The combined effect of these two factors can be seen in the velocity-vorticity temporal correlation where after an initial dip, the correlation function shows a saturation at larger time scales (for all q-values other than $q=1$). At $q=1$, while the dip is even more pronounced, the profile shows large wavelength Alfven waves \cite{chattopadhyay2013} with a saturating envelope for the correlation function. Similar trends could be seen in connection with all temporal correlation functions, and hence for the ensemble averaged energy growth too. The fact that there is an additional energy input associated with a nonzero noise cross-correlation comes out across all these correlation functions in that the relative magnitudes are always one to two orders of magnitude higher than their equivalent zero cross-correlated noise cases \cite{chattopadhyay2013,nath2013}. The qualitative summary for the temporal autocorrelation functions of the different variables (Figs 1-4) too portray very similar patterns. Apart from an overall increase in transient energy, the same saturation tendency (all q-values) with large wavelength oscillations for $q=1$ are omnipresent in all of velocity, vorticity and magnetic vorticity related autocorrelations. The magnetic field autocorrelation is a special and most reassuring case in that profiles for all q-values now show oscillating Alfven waves that could be easily attributed to an effective increase in the magnetic field energy due to this time symmetry violating effect. A remarkable difference with the nonzero cross-correlation could also be seen from the fact that the profiles for the different q-values are now distinct. This can be visualized as a classical analogue of the Zeeman effect where energy degeneracy is removed through additional magnetic energy influx in the system. 

Compared to the temporal correlations (both autocorrelation and cross-correlations), the cases for the spatial correlations are rather \enquote{less interesting} in that the results are obvious and analogous to their zero-noise-cross-correlated counterparts. Apart from quantitative differences in energy magnitudes, the reason for which has already been explained, the decaying profiles reassert the initial hypothesis of nonzero noise cross-correlation as being predominantly a time symmetry effect. To summarize, a nonzero noise cross-correlation (and hence time symmetry violation too) boosts the energy levels in the system while simultaneously neutralizing the oscillating effects generally attributed to Alfven waves whose major effects could be seen in all forms of temporal statistics.

Many issues in relation to such subcritical (linear) turbulence in rotating accretion flows still remain unclear. Although it is now rather well acknowledged that transient linear turbulence is essentially caused by non-normal disturbances \cite{man2005,maretzke2014} (indirectly shown in \cite{chattopadhyay2013} too), the implication of the resultant power law (energy growth rate $\sim~{\text{Re}}^{2/3}$) with reference to stochasticity is a question that is pretty much unresolved as yet. Also, how all of these patterns evolve and modify the dynamic bifurcation structures in relevant phase diagrams \cite{man2005}, especially in the non-equilibrium nonlinear regime, are exciting questions that need to be delved with. Through this paper what we have now shown is that a nonzero noise cross-correlation is effectively another independent (field theoretically) \enquote{relevant} variable whose effects are expected to affect all such situations through time symmetry violations that needs to be appropriately dealt with in all accretion flow related situations.

\section*{Acknowledgments}
Both authors acknowledge discussions with Banibrata Mukhopadhyay. AKC also thanks the Royal Society, U.K., research grant number RG110622, for partial support.

{}

\end{document}


\title
{Supplementary information for the manuscript: \textquotedblleft Cross-correlation Aided Transport in Stochastically Driven Accretion Flows\textquotedblright}

\author{Sujit Kumar Nath}
\affiliation{Department of Physics, Indian Institute of Science, Bangalore 560 012, India}
\email{sujitkumar@physics.iisc.ernet.in}

\author{Amit K Chattopadhyay}
\affiliation{Aston University, Nonlinearity and Complexity Research Group, Engineering 
and Applied Science, Birmingham B4 7ET, UK}
\email{a.k.chattopadhyay@aston.ac.uk}

\maketitle

\section{Supplementary Section}

A non-zero noise cross-correlation implies a violation of time translation symmetry making it imperative that only temporal correlation functions, both for autocorrelation and cross-correlations, should be the primarily affected ones. This is what we find too as have been explained in details in the main text (article). This noise cross-correlation brings about some perceptible quantitative changes in the spatial correlation functions too. Although there is not much change in the spatial physics part, in view of the fact that quantitative changes still occur, here we present all other spatial correlation plots which are not presented in the main text for the convenience of the more interested readers. The implication of the q-values remains the same as in the main text: $q=3/2$ represents a Keplerian disk, $q=2$ represents a disk with a constant angular momentum while $q=1$ represents a disk with a flat rotation curve.

\begin{figure}[H]
 \centering
\includegraphics[scale=0.9]{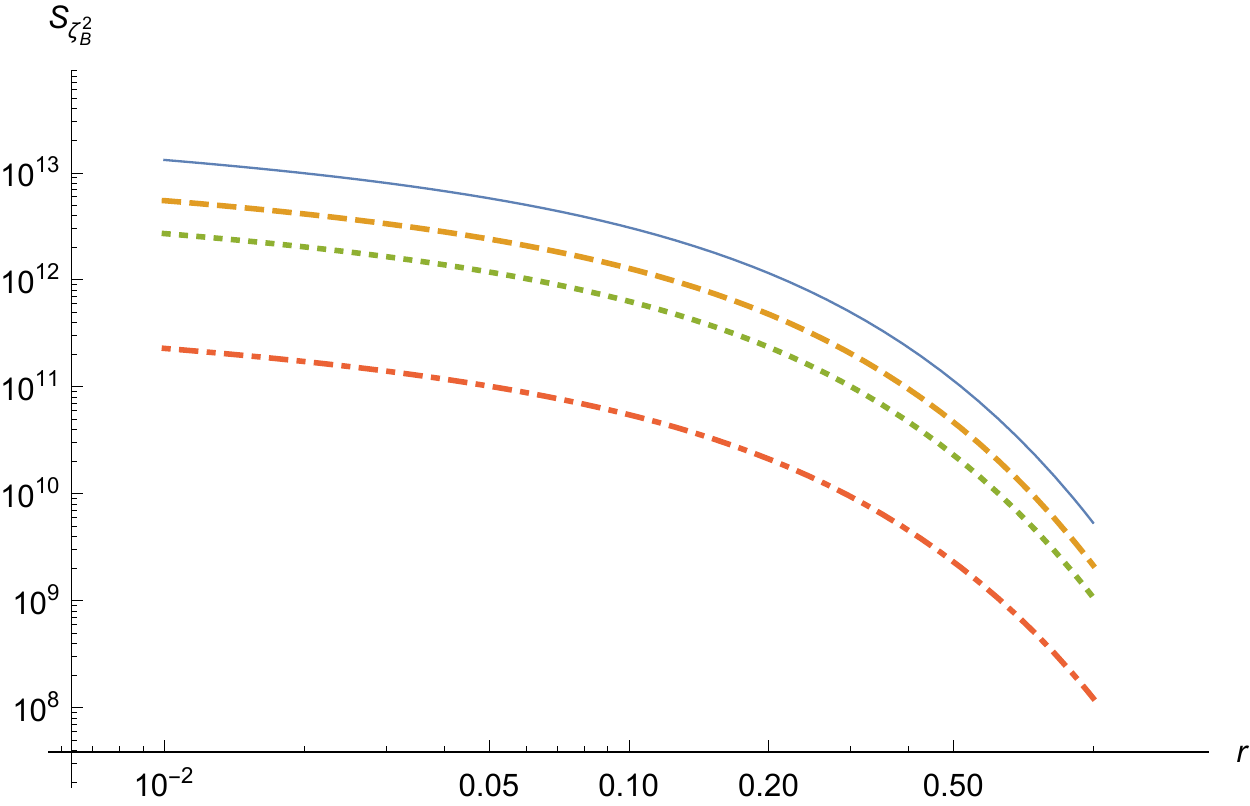}
\caption{(Color online) Spatial autocorrelations of magnetic vorticity ($S_{\zeta_B^2}$) upto $r=1.0$.  The solid line represents $q \approx 2$, the dashed line is for $q=1.7$, the dotted line represents $q=1.5$ and the dot-dashed line is for $q=1$.}
\label{zbzb_spatial}
\end{figure}

\begin{figure}[H]
 \centering
\includegraphics[scale=0.9]{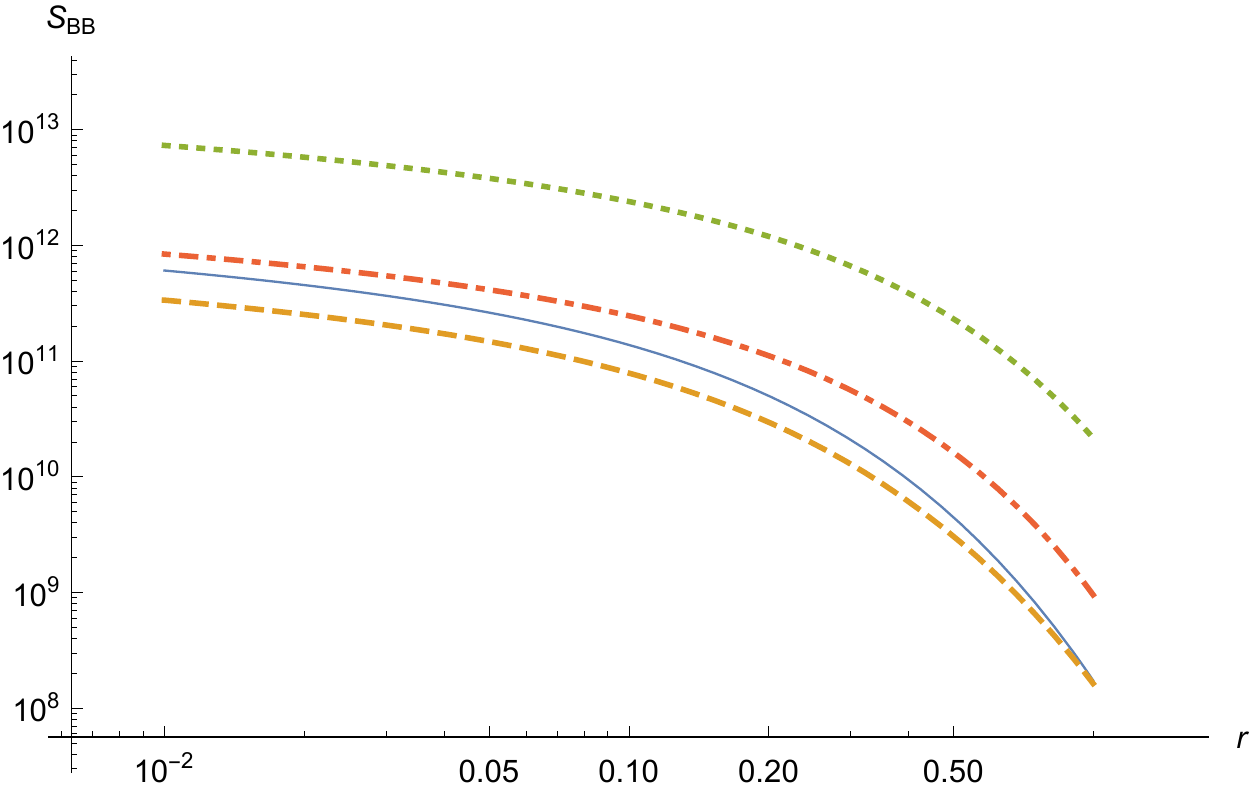}
\caption{(Color online) Spatial autocorrelations of magnetic field ($S_{BB}$) upto $r=1.0$. Plot styles are same as in Fig~\ref{zbzb_spatial}.}
\label{bb_spatial}
\end{figure}

\begin{figure}[H]
 \centering
\includegraphics[scale=0.9]{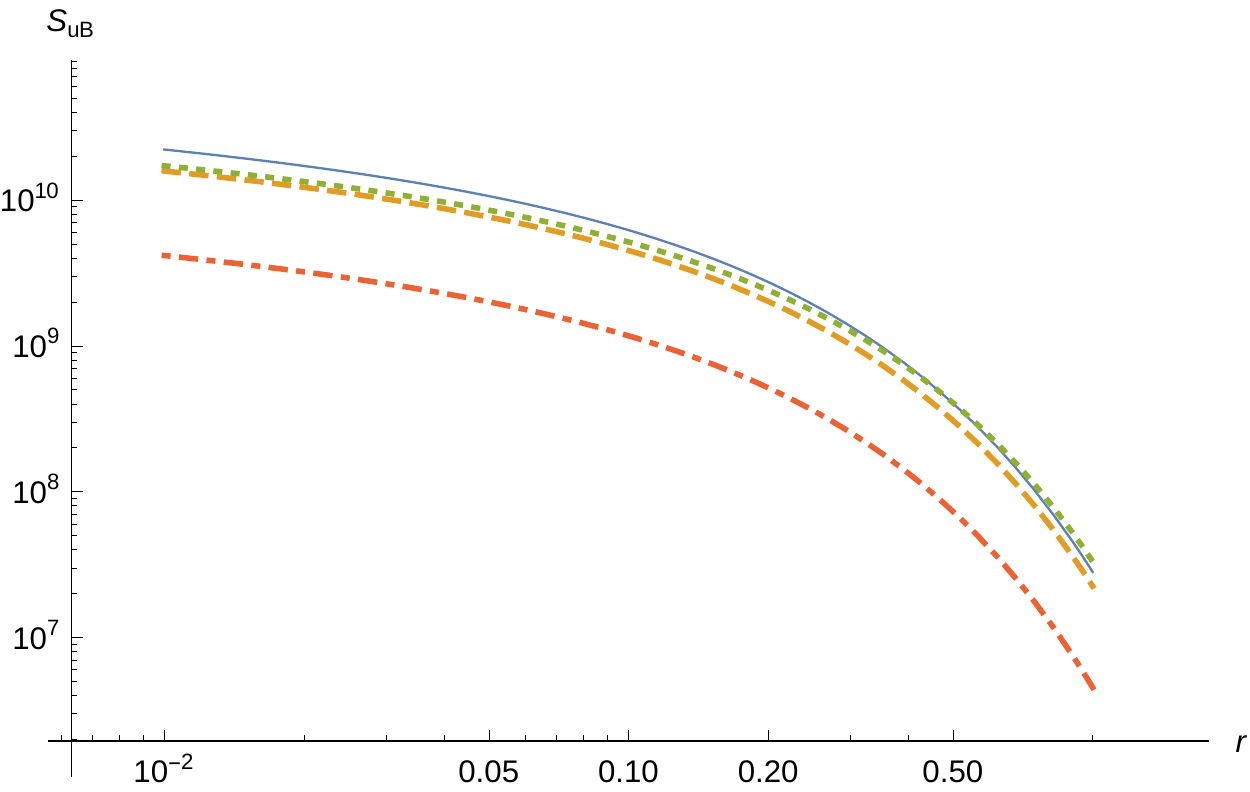}
\caption{(Color online) Spatial correlations of velocity and magnetic field ($S_{uB}$) upto $r=1.0$. Plot styles are same as in Fig~\ref{zbzb_spatial}.}
\label{ub_spatial}
\end{figure}

\begin{figure}[H]
 \centering
\includegraphics[scale=0.9]{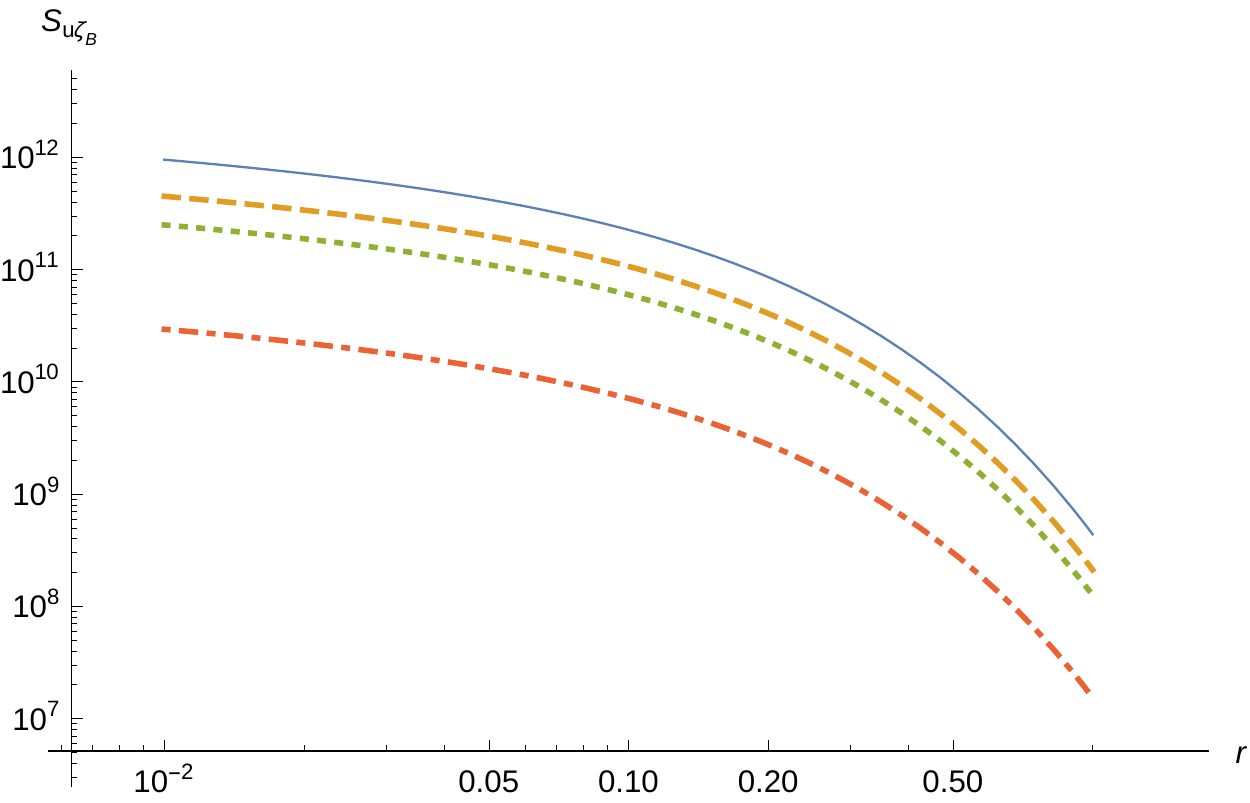}
\caption{(Color online) Spatial correlations of velocity and magnetic vorticity ($S_{u\zeta_B}$) upto $r=1.0$. Almost no difference in magnitude compared to the zero noise cross-correlation case. Plot styles are same as in Fig~\ref{zbzb_spatial}.}
\label{uzb_spatial}
\end{figure}

\begin{figure}[H]
 \centering
\includegraphics[scale=0.9]{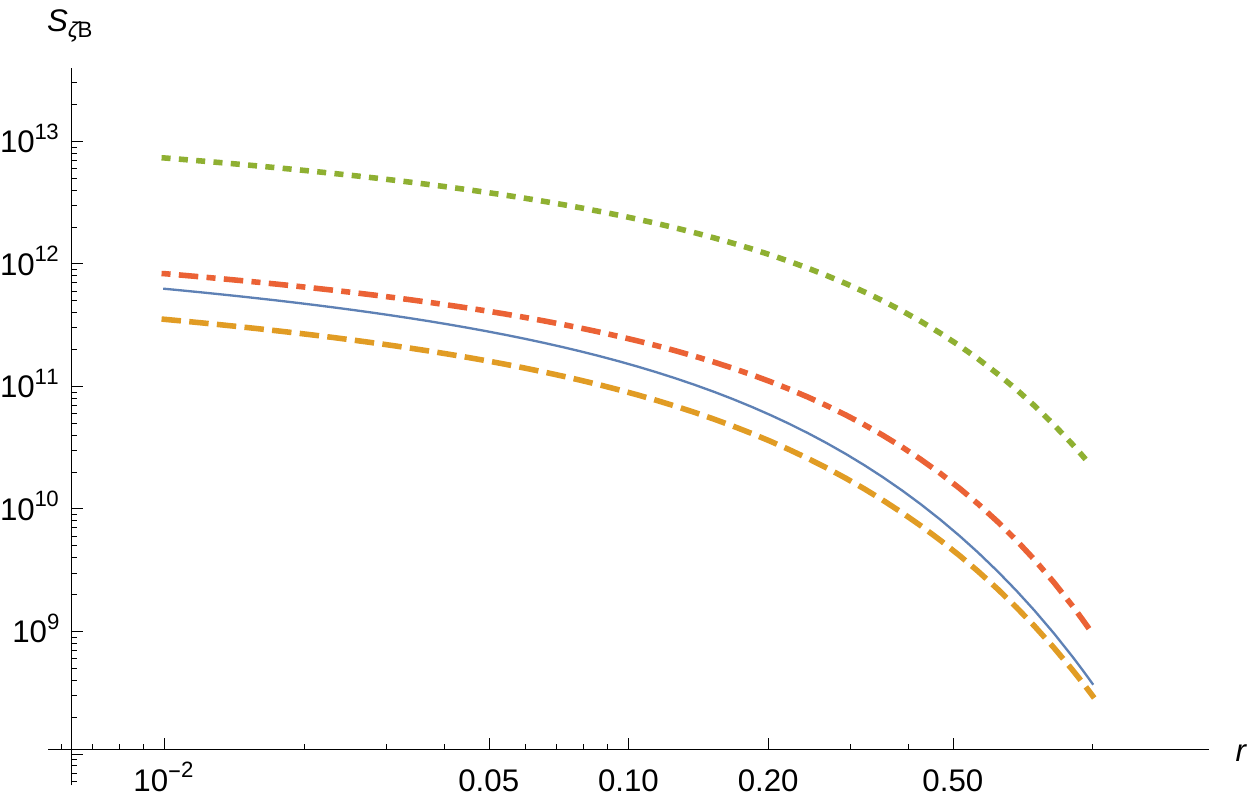}
\caption{(Color online) Spatial correlations of vorticity and magnetic field ($S_{\zeta B}$) upto $r=1.0$. Similar in magnitude compared to the zero noise cross-correlation case. Plot styles are same as in Fig~\ref{zbzb_spatial}.}
\label{zb_spatial}
\end{figure}

\begin{figure}[H]
 \centering
\includegraphics[scale=0.9]{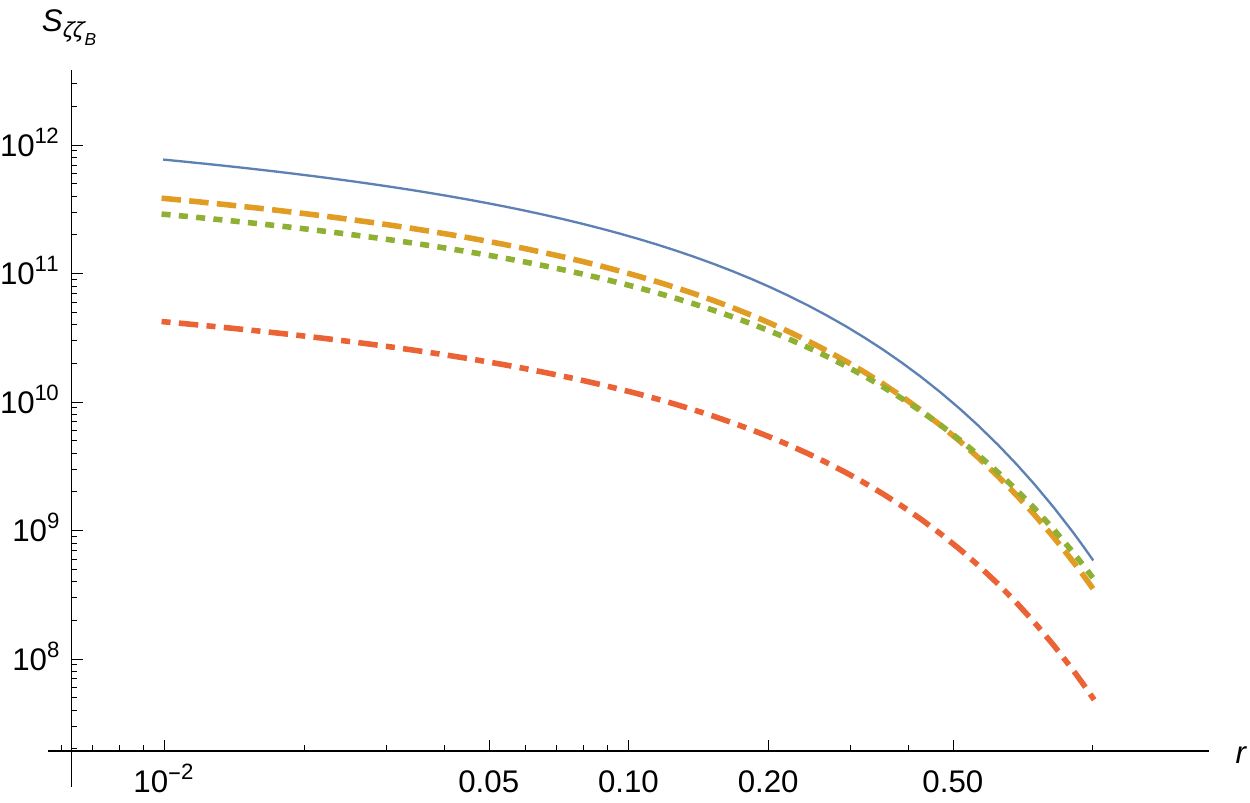}
\caption{(Color online) Spatial correlations of vorticity and magnetic vorticity ($S_{\zeta \zeta_B}$) upto $r=1.0$. 8 orders of magnitude higher than the zero noise cross-correlation case. Plot styles are same as in Fig~\ref{zbzb_spatial}.}
\label{zzb_spatial}
\end{figure}

\begin{figure}[H]
 \centering
\includegraphics[scale=0.9]{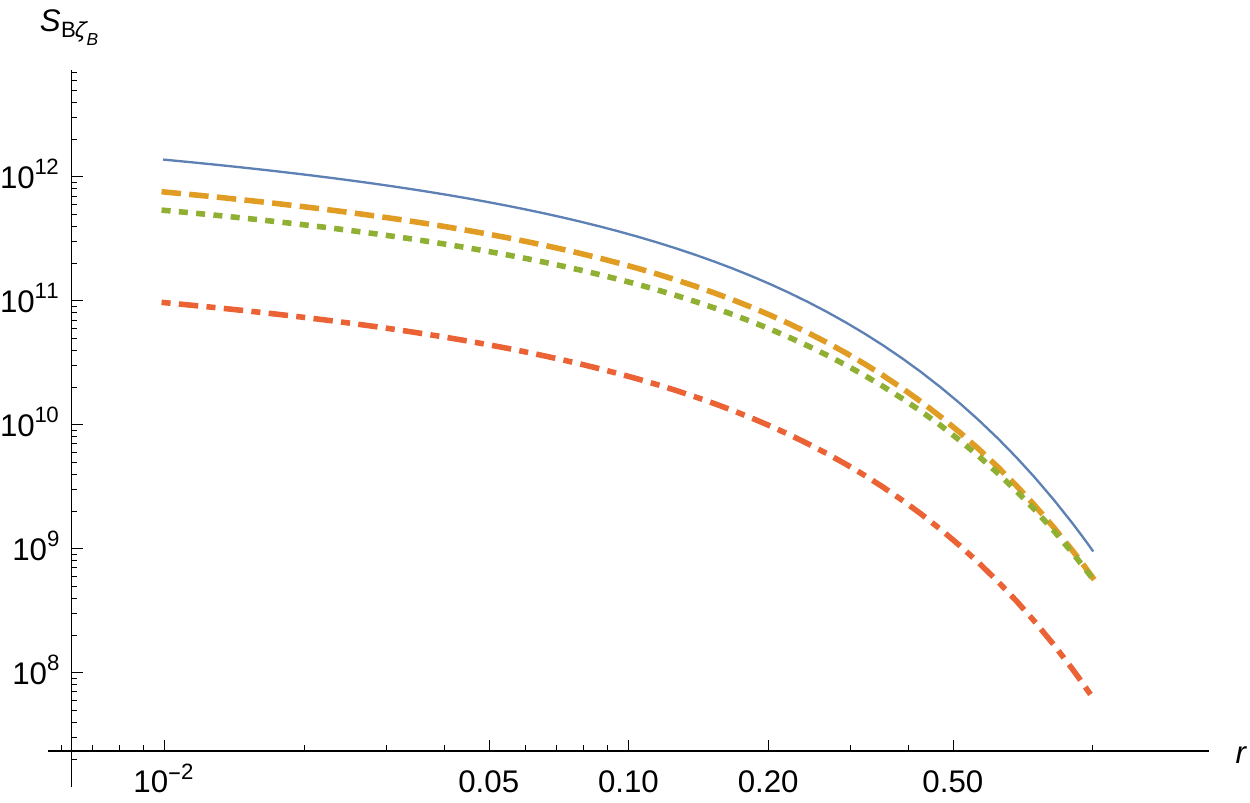}
\caption{(Color online) Spatial correlations of magnetic field and magnetic vorticity ($S_{B\zeta_B}$) upto $r=1.0$. 8 order of magnitude higher than the zero noise cross-correlation case. Plot styles are same as in Fig~\ref{zbzb_spatial}.}
\label{bzb_spatial}
\end{figure}

\begin{figure}[H]
 \centering
\includegraphics[scale=0.9]{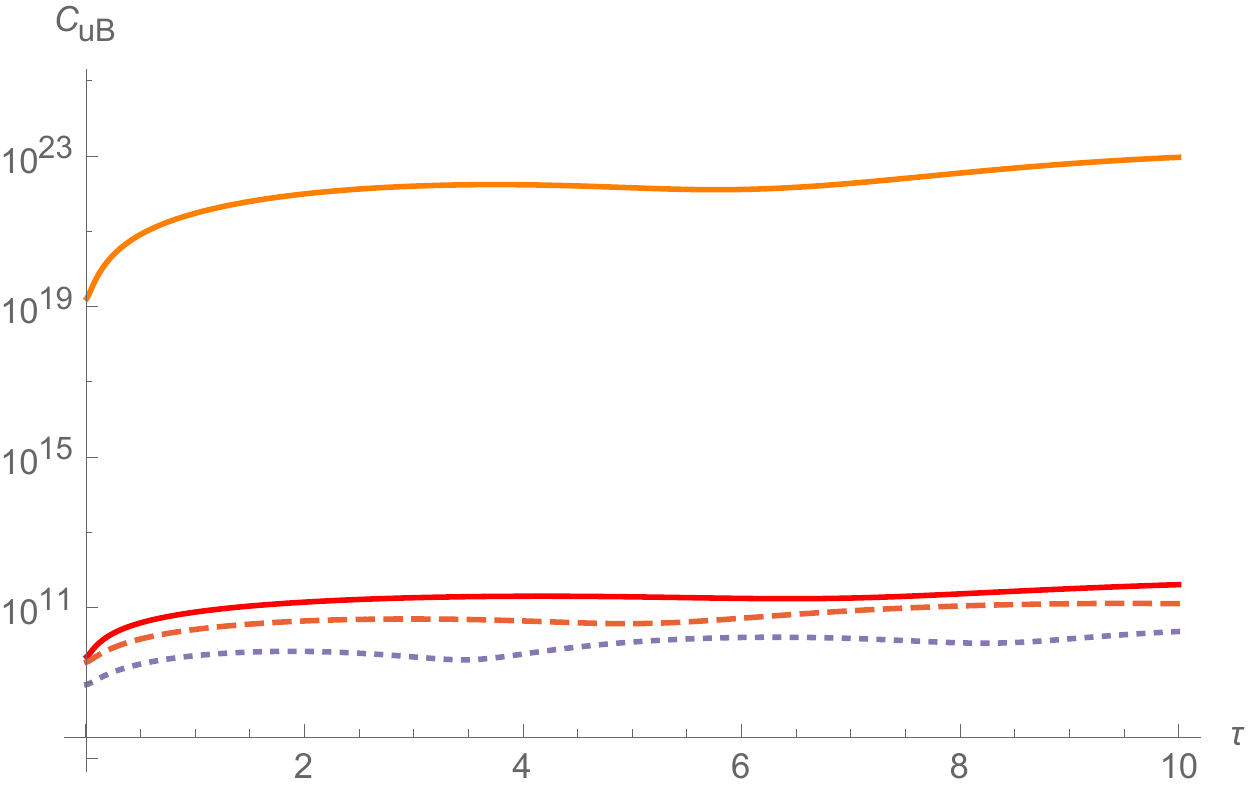}
\caption{(Color online) Temporal cross-correlation of the velocity with the magnetic field ($C_{uB}$) plotted against time separation $\tau$. This plot shows the effect of the limit $q \to 2$ as $q$ becomes increasingly closer to 2. The thick line shows the result for $q=1.9999$, the red line represents $q=1.99999$ while the orange line shows $q=1.999999$. The dashed line is for $q=1.7$, the dotted line represents $q=1.5$ and the dot-dashed line is for $q=1$. Clearly, a progressive increase in the q-value approaching 2 only causes a quantitative shift in the correlation function while keeping the qualitative result (Alfven oscillation) unchanged. In the plot shown, the $q=1.9999$ and $q=1.99999$ are almost indistinguishable because they merge in this limit.}
\label{ub_temporal}
\end{figure}
